\documentclass[%
 reprint,
 amsmath,amssymb,
 aps,
prb,
]{revtex4-2}

\usepackage{graphicx}
\usepackage{dcolumn}
\usepackage{bm}

\usepackage{graphicx}
\usepackage{subfigure}
\usepackage{booktabs}
\usepackage{amssymb,amsmath}
\usepackage{color}
\usepackage[english]{babel}
\usepackage[utf8]{inputenc}
\usepackage[colorinlistoftodos, color=green!40, prependcaption]{todonotes}
\usepackage{amsthm}
\usepackage{mathtools}
\usepackage{physics}
\usepackage[dvipsnames]{xcolor}
\usepackage{graphicx}
\usepackage{adjustbox}
\usepackage{placeins}
\usepackage{csquotes}
\usepackage{bm}
\usepackage{lipsum}
\usepackage[normalem]{ulem}
\usepackage{amssymb}
\def\rbs{\boldsymbol{r}}
\DeclareMathOperator{\BesselY}{Y}
\DeclareMathOperator{\StruveH}{H}

\colorlet{linkColour}{magenta}
\colorlet{citeColour}{OliveGreen}
\colorlet{urlColour}{cyan}

\usepackage[colorlinks=true]{hyperref}
\hypersetup{
    unicode=false,          
    pdftoolbar=true,        
    pdfmenubar=true,        
    pdffitwindow=false,     
    pdfstartview={FitH},    
    pdftitle={Fractional trions},    
    pdfauthor={Hart Goldman},     
    pdfsubject={Subject},   
    pdfcreator={Hart Goldman},   
    pdfproducer={}, 
    pdfkeywords={keyword1} {key2} {key3}, 
    pdfnewwindow=true,      
    colorlinks=true,       
    linkcolor=linkColour, 
    citecolor=citeColour,        
    filecolor=linkColour,      
    urlcolor=urlColour           
}

\usepackage{comment}

\begin{document}

\title{Anyonic molecules through the trion looking glass}

\author{Hao-Ran Cui$^1$, Vladimir Calvera$^{1,2}$, Umang Mehta$^1$, and Hart Goldman$^1$\\
$^1$ \emph{School of Physics and Astronomy, University of Minnesota, Minneapolis, MN 55455, USA }\\
$^2$ \emph{William I. Fine Theoretical Physics Institute, University of Minnesota, Minneapolis, MN 55455, USA}}

\date{\today} 

\begin{abstract}
Motivated by recent optical measurements in twisted MoTe$_2$ homobilayers, we numerically study the binding of itinerant anyons to ordinary electronic bound states. Using a minimal few-body model of an anyon alongside holes and electrons in a Gaussian trapping potential, we show that a fractionally charged bound state comprised of an anyon and an ordinary electronic trion can become favorable nearby local electrostatic defects. The resulting object is an anyonic molecule wherein the anyon is bound to the trion ``nucleus,'' which in turn is localized to the defect and can be excited optically. We find that the resulting anyon-trion can exhibit a photoluminescence red-shift relative to the ordinary trapped trion, with binding on the meV scale for parameters relevant to twisted MoTe$_2$. We find that the value of this red-shift depends on \emph{both} the anyon's fractional charge and its kinetic energy, thereby encoding both universal and non-universal anyon properties. Our results are qualitatively consistent with observations in twisted MoTe$_2$.

\end{abstract}

\maketitle

\section{\label{sec:intro}Introduction}

Two-dimensional moiré materials exhibiting the fractional quantum anomalous Hall (FQAH) effect~\cite{cai2023signatures,zeng2023thermodynamic,xu2023observation,park2023observation,Li2025} offer unprecedented opportunities to study and control fractionally charged quasiparticles -- anyons -- in a highly tunable environment. In these new platforms, bulk anyons can hop along the moiré superlattice and acquire dispersion \cite{tang2013superconductivity,kim2025topological,schleith2025anyon,yan2025anyon,shi2025doping,divic2025anyon,shi2025dopingB,pichler2025microscopic,gonccalves2025spinless,Iyer2026}, in contrast to traditional quantum Hall settings where their kinetic energy is quenched by the strong perpendicular magnetic field. We are thus entering an era where \emph{anyon dynamics} can be studied experimentally, with potential for design of correlated multi-anyon systems. 

One emerging avenue is the optical manipulation of anyons localized to defects~\cite{rashba1993anyon,Wojs2000,Wagner:2025kig,paul2025shininglightcollectivemodes,mostaan2025anyon}. In transition metal dichalcogenide (TMD) materials like twisted MoTe$_2$ ($t$MoTe$_2$), nano-scale impurities are already known to trap excitonic complexes~\cite{He:2014bnl,srivastava2015optically,Chakraborty2015,koperski2015single,Kumar_2015,thureja2023electricallytunablequantumconfinement,li2026signatures}, which can be excited with light. The energy scales of these bound states are thus visible in photoluminescence (PL) measurements. If the surrounding correlated electron fluid is fractionalized, anyons in the vicinity of the trap may become ensnared alongside ordinary electrons and holes, leading to fractionally charged \emph{anyonic molecules}. These bound objects are combinations of one or more anyons with an exciton or trion, and their properties thus depend on (1) the anyons' universal topological data -- their fractional charge and braiding statistics -- and (2) their non-universal dynamical properties, such as their bandwidth. Probing anyonic molecules therefore provides a crucial observable window into anyon dynamics.

\begin{figure}[t]
    \centering
    \includegraphics[width=0.98\linewidth]{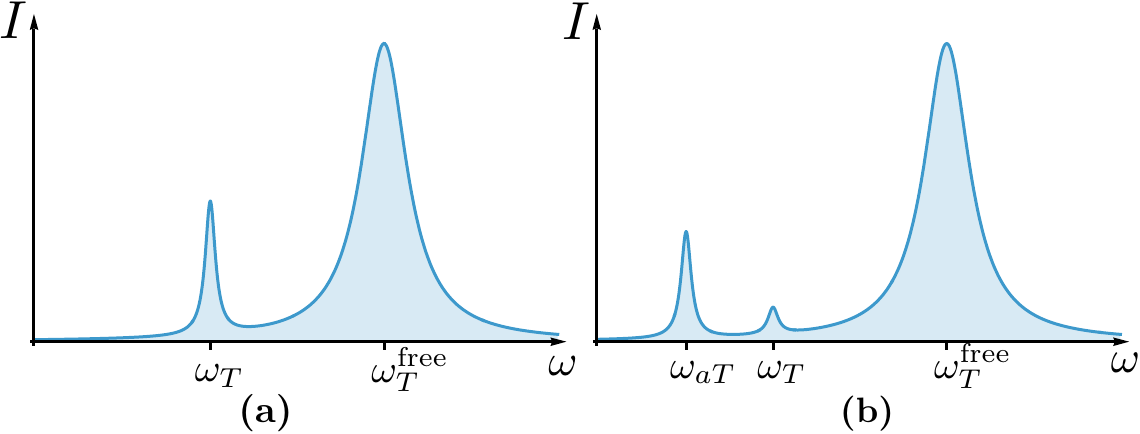}\\\vspace{10pt}
    \includegraphics[width=0.45\textwidth]{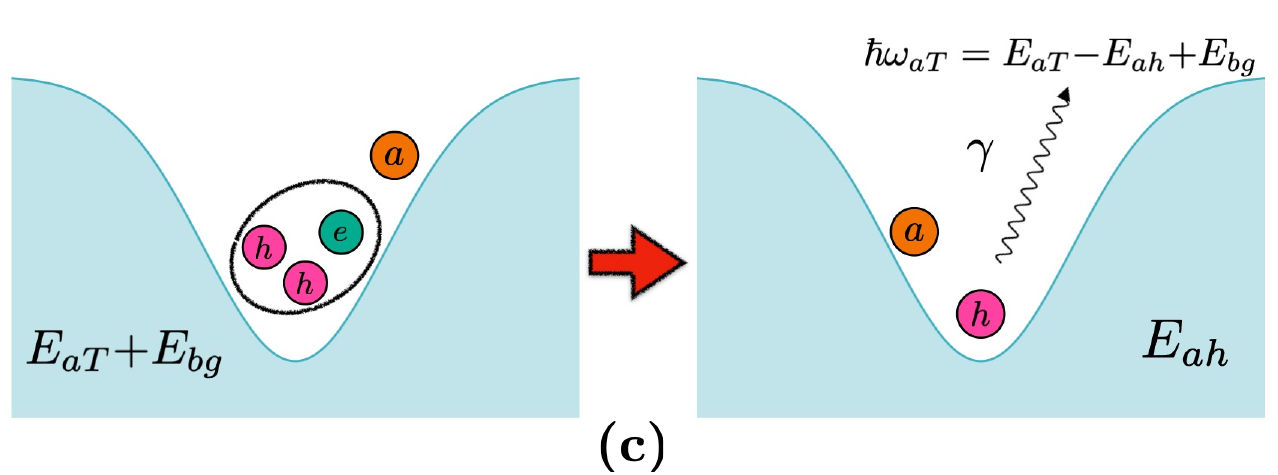}
    \caption{Sketch of the PL intensity seen at defect `traps' in Ref.~\cite{Li2025}. (a) Deep in the FQAH plateau, a trapped trion peak is seen at a red-shifted frequency $\omega_T$ relative to the un-trapped trion peak  at $\omega_T^{\mathrm{free}}$.  (b) On doping, an additional peak is observed exhibiting a further red-shift. (c) We model the corresponding bound state as an \emph{anyonic molecule} localized to a potential trap.  This exotic object is made up of an anyon  attached to a trion --two holes and an electron-- with energy $E_{aT}$. When the exciton recombines, the anyon and remaining hole are deposited back into the trap region with energy $E_{ah}$, and a photon of frequency $\omega_{aT}=E_{aT}-E_{ah}+E_{bg}$ is emitted, where $E_{bg}$ is the band gap. }
    \label{fig:PLSketch}
\end{figure}

Anyonic molecules of this type may already have been observed. Recently, unusual new PL features were measured in the FQAH regime of hole-doped $t$MoTe$_2$~\cite{li2026signatures}, nearby defects hosting trapped trions consisting of two holes and one electron. On hole doping the system away from the centers of the $\nu=2/3$ and $\nu=3/5$ FQAH plateaus, a red-shift ${\Delta \omega\sim\mathcal{O}(1\,\mathrm{meV})}$ in the trion spectral peak was seen consistently across several defects (see Fig.~\ref{fig:PLSketch} for a sketch). The close proximity of this feature to the undoped trion spectral peak suggests a weak deformation of the trion wave function. 

The most tantalizing explanation is the binding of the trion to an itinerant anyon of minimal charge ($e/3$ or $e/5$), with the red-shift corresponding to a 
loss of Coulomb energy. Interestingly, the authors of Ref.~\cite{li2026signatures} reported that the ratio of the red-shift magnitudes between the two plateaus, ${\Delta \omega_{2/3}/\Delta \omega_{3/5}\approx (e/3)/(e/5)=5/3}$, appeared quantized across multiple defects, suggesting a connection between the PL red-shift and the binding energy of a fractionally charged particle.

Nevertheless, that a trion can \emph{lower} its energy by plucking an anyon out of the ambient fractionalized fluid is far from obvious: Both the anyon and the trion carry positive charge and repel each other! Therefore, a more detailed quantum mechanical justification for the formation of such an anyon-trion complex is necessary.

In this work, we develop a simple few-body quantum mechanics model where the formation of anyon-trion composites can be studied  using variational techniques. We start with a deep Gaussian potential well on the scale of the moiré unit cell, which we postulate 
traps 
an (un-fractionalized) hole (Fig.~\ref{fig:PLSketch}). Shining light on the resulting defect can excite a virtual exciton, leading to a trapped trion. Doping charge into the system introduces a population of itinerant anyons, which can in turn become localized to the defect region and bind to trapped trions, forming an anyonic molecule. By computing the binding energies at each stage, we can obtain predictions for observable PL spectra. 

Surprisingly, we find that a \emph{red-shifted} anyonic molecule is possible due to an interplay of electrostatic and quantum effects. In the presence of the trapping potential, a trion may indeed lower its energy by binding to an itinerant anyon, whose wave function is maximized away from the trap center. When this anyonic molecule decays, the anyon and a hole relax into the region of the trap, and a photon is emitted. This photon's frequency is the energy difference between these two initial and final states (Fig.~\ref{fig:PLSketch}). In a realistic parameter regime we find that this photon is red-shifted -- in qualitative agreement with the experiment of Ref.~\cite{li2026signatures} -- although a blue-shift is also possible (see Fig.~\ref{fig:V0_vs_sigma_stability}). Furthermore, we find an intriguing sensitivity to anyon dynamics: For a red-shift to occur in our model, the anyon kinetic energy cannot be too small. 

Our phenomenological model supports  trion PL as an tool for detecting bound anyons and studying their dynamics. Because our framework incorporates the intricate electrostatic screening effects stabilizing the trion as well as the anyon kinetic energy, we are able to assess the energetic competition of anyonic molecules with  ordinary trions and compare against experiments. 

We proceed as follows. In Sec.~\ref{sec:model}, we introduce the model. In Sec.~\ref{sec:quantum chemistry}, we describe our method and report results in the absence of a trap. In Sec.~\ref{sec: pl red-shift}, we calculate the photon frequency shift due to anyon binding in the presence of a trap. In Sec.~\ref{sec:fractional charge}, we study how the latter shift depends on anyon properties. In Sec.~\ref{sec:discussion} we summarize our results.

\section{\label{sec:model}Model}
Our goal is to study the formation of anyonic bound states near defects. To this end, we assume that the energetics of anyonic molecules -- as well as ordinary excitons and trions -- can be captured by a ``few-body'' quantum mechanics model. We introduce a Hamiltonian describing electrons, holes, and (at most) one anyon in the presence of a trapping potential,
\begin{equation}\label{eq:Ham1}
    H = \sum_{j} \left(\frac{\boldsymbol{p}_j^2}{2m_j} - q_j U(\boldsymbol{r}_j) \right)+ \sum_{i<j} q_i q_j V(\abs{\boldsymbol{r}_i-\boldsymbol{r}_j})\,.
\end{equation}
Here $i,j$ are particle labels, which run over electrons~($e$), holes~($h$), and the anyon~($a$); $\boldsymbol{r}_j$ and $\boldsymbol{p}_j$ are two-dimensional position and momentum operators of particle $j$; $q_j$ and $m_j>0$ are the charge (in units of the electron charge) and the mass of particle $j$. The latter is phenomenological input.
In particular, we work with an effective model of quadratically dispersing anyons, whose effective mass is an unknown parameter; recent work \cite{yan2025anyon,gonccalves2025spinless} has indicated anyon bandwidths significantly narrower than those of the holes and electrons. As such, our modelling of anyons does not assume projection into a Chern band or Landau level. Our convention for charge assignments is $q_h =1$, $q_e = -1$, and $q_a>0$ is a rational fraction, {i.e.} we assume that anyons are quasi-holes. Hence the trap repels electrons and attracts holes and anyons. 

We model electrostatic interactions between particles in the dielectric environment of $t$MoTe$_2$ using the Keldysh-Rytova interaction potential~\cite{rytova2018screened,Keldysh1979,cudazzo2011dielectric}, $V(r)$,
\begin{equation}\label{eq:VRK:Ineraction}
    V(r) =  \frac{\pi e^2}{2\kappa\, r_0}\frac{1}{4\pi\varepsilon_0}\Big[\StruveH_0(r/r_0)- \BesselY_0\left(r/r_0\right)\Big]\,.
\end{equation}
Here $\StruveH_0$ and $\BesselY_0$ are the zeroth Struve H function and the Bessel function of second kind, respectively. The material-dependent parameters $r_0$ and $\kappa$ are respectively the effective screening length and the relative dielectric constant. This interaction interpolates between a Coulomb potential, ${V(r) \sim 1/r}$, at large distances, ${r\gg r_0}$, and a logarithmic interaction, ${V(r) \sim \frac{1}{r_0}\log(r_0/r)}$, at short distances, ${r \ll r_0}$. 

We describe the effect of the defect through a one-body scalar potential, $U(\boldsymbol{r})$, with Gaussian profile, 
\begin{equation}
    U(\boldsymbol{r}) = U_0 \exp(- \frac{\abs{\boldsymbol{r}}^2}{2\sigma^2}).
\end{equation}
Here $U_0 \geq 0$ is the overall magnitude of the trapping potential, and $\sigma$ is a measure of its size. 
We will see that our results are robust over a physically relevant range of parameter choices. 

We assume that prior to doping anyons into the system, this potential well traps a single hole. Shining light on the resulting positively charged defect can then nucleate an additional exciton, which binds to the hole to form a trapped trion. Although we find this assumption physically reasonable, we emphasize that the microscopic nature of the defects in $t$MoTe$_2$ is poorly understood. More elaborate models for the defect may be relevant to experiments and may possess quantitative differences with our minimal model. 

To study problems with anyons, one must generally include an emergent gauge field implementing fractional exchange statistics. However, the electrons and holes in our model are local particles, meaning that they are neutral under statistical gauge fluctuations. Hence, interactions of a single anyon with ordinary electrons and holes should be purely electrostatic~\footnote{Although one anyon is not a strictly speaking gauge invariant object in itself, we will assume that other anyons are present in the sample such that the many-body wave function of the whole system is gauge invariant}.
Nonetheless, possibilities for charged multi-anyon molecules stabilized by the correlated FQ(A)H environment (without trapping potentials)~\cite{Munoz2020,gattu2025molecular,Yang2025}, as well as for fractional excitons~\cite{LAUGHLIN1984,Haldane1985,Kamilla1996,Yang2012,PhysRevB.95.075201}, have also attracted great recent interest. In both cases, a careful study of the interplay of braiding with electrostatic interactions is essential. We studied one such example -- a bound state between a quasi-hole and quasi-particle -- in App.~\ref{app: fractional exciton energies}.

\section{\label{sec:quantum chemistry}Fractionalized quantum chemistry}

Solving the few-body Schr\"odinger equation directly is difficult in practice. We thus employ a technique commonly used in chemistry; namely, the `Correlated Gaussian method' \cite{suzuki2002stochastic,mitroy2013theory}. This method has been previously used to study trions and excitons in a wide range of systems~\cite{katow2017numerical,van2017excitons,van2018excitons,cho2021simulations,tenorio2026gaussian}.

To construct the variational ground state wavefunction, we choose a basis of explicitly correlated Gaussian functions (ECG)~\cite{mitroy2013theory}, 
\begin{align}
\phi_{\alpha}=\exp\Big(-\sum_{i,j=1}^{N}M^\alpha_{ij}\,\boldsymbol{r}_i\cdot \boldsymbol{r}_j\Big),
\label{eq: minimal ECG basis}
\end{align}
labeled by $\alpha$, where $M^\alpha_{ij}$ are the elements of a $N\times N$ dimensional matrix with $N$ being the particle number. The correlation between different particles is built explicitly in the off-diagonal components of $M^{\alpha}_{ij}$. A key advantage of the Gaussian basis is that it provides fast numerical convergence and analytical tractability of the matrix elements (see App.~\ref{app:Numerical Methods}). Note that we restrict our calculations to states with vanishing total angular momentum.

We construct a variational ground state wavefunction as $\psi=\sum_{\alpha}C_{\alpha}\phi_{\alpha}$. The ground state energy, $E_{\text{min}}$, is then the solution to the generalized eigenvalue problem,
\begin{align}
   \sum_{\beta} \langle \phi_{\alpha}|H|\phi_{\beta}\rangle\, C_{\beta}=E_{\text{min}}\sum_{\beta}\langle \phi_{\alpha}|\phi_{\beta}\rangle\, C_{\beta}\,,
\end{align}
with $C_{\alpha}$ optimized for the smallest possible $E_{\text{min}}$. We carry out this optimization using the stochastic variational method, which randomly proposes a basis element, $\alpha$, which is accepted only if it decreases the energy (see App.~\ref{app:Numerical Methods} for details). 

In our calculations, we adopt parameter choices relevant to MoTe$_2$. 
We set the electron and hole masses to their monolayer values,  $m_e=m_h=0.65\, m_0$, where $m_0$ is the bare electron mass. We believe these values to be especially relevant when the electron and hole wave functions have much narrower spatial extent than the unit cell size of $t$MoTe$_2$, roughly $6\,\rm{nm}$ at the experimentally relevant twist angle of $3.3^\circ$. Choosing different values of the electron and hole masses -- particularly for the hole that is drawn from the moir\'{e} mini-band -- does not dramatically change our qualitative results, although it does reduce the quality of our numerical data. 
Additionally, we adopt interaction parameters ${\kappa = 4}$ and ${r_0 = 3.17\,\rm{nm}}$.

With these parameters, we begin by checking the binding energies of the exciton ($X$) and trion ($T$). These objects respectively correspond to the ground states of the electron-hole problem and the electron-two holes problem. In the latter case, we always assume the holes' wave function is a spin singlet that is even under exchange of their spatial coordinates.  

In the ``free'' situation without a trapping potential, ${U_0=0}$, we obtain the exciton and trion binding energies, 
\begin{align}
{\delta}E_X^{\rm{free}}\approx 190.47\,\mathrm{meV}\,,\qquad {\delta}E_T^{\mathrm{free}}\approx 14.59\,\rm{meV}\,,
\end{align}
with numerical error of $\mathcal{O}(10^{-3}\,\mathrm{meV})$. These values are comparable to the measured free exciton and trion binding energies in monolayer MoTe$_2$, and they are of the same order of magnitude as the values reported in $t$MoTe$_2$ at charge neutrality~\cite{cai2023signatures,li2026signatures}. However, we emphasize that in moir\'{e} $t$MoTe$_2$ all such energy scales will evolve smoothly with doping and will depend on the strongly correlated environment in the FQAH regime. We furthermore compute the root-mean-squared (RMS) size of the trion wave function to be ${\sqrt{r_T^2}\approx 2.33\,\mathrm{nm}}$, about a third of the size of the moiré unit cell~\footnote{We used the definition $r_T^2=\frac{1}{3}\expval{(r_{h_1h_2}^2+r_{eh_1}^2+r_{eh_2}^2)}$.}.

Having benchmarked our calculations against the ordinary exciton and trion, we introduce an itinerant anyon with charge $q_a$ and mass $m_a$. Our primary interest is whether this anyon can bind to the trion comprised of two holes and one electron. We again start with the ``free'' case with the trapping potential is turned off, $U_0=0$. In hole-doped FQAH phases, the itinerant anyons will have charge of the same sign as the holes, meaning that they should be repelled by the trion. This expectation is consistent with our numerical calculations: We searched for such anyonic molecule bound states for a wide range of $q_a, m_a$ values, also sweeping values of $\kappa$, $r_0$, and the electron and hole masses. In all cases, we found that the anyonic molecule either dissociates into a trion alongside a decoupled anyon, or a more exotic scenario where the anyon binds to an exciton to form a bound state, leaving a decoupled hole (see App.~\ref{app: AdditionalData}). The absence of a free anyon-trion bound state is consistent with experiment, since the proposed anyon-trion PL peaks exclusively appear in regions of the sample hosting a trapped trion.

We remark that the fractionally charged trion state composed of an electron, a hole, and an anyon has binding energy much smaller than $\delta E_T^{\mathrm{free}}$ for minimally charged anyons. Interestingly, we find that for anyon charges $q_a\gtrsim 2/3$, the formation of the fractionally charged trion can become more favorable than the free trion. The separation of binding energies becomes especially noticeable as anyon mass decreases.
This result suggests that additional PL peaks could appear if the itinerant anyon carriers fuse into higher-charge composites, e.g. $2e/3$, themselves. 

Anyonic molecule bound states become much more favorable in the environment of the trapping potential, $U$. If the potential strength $U_0>0$, the defect will attract the holes and the anyon, while repelling the electron. The trap thus aids the stability of the molecule, by counterbalancing the mutual repulsion between the holes and the anyon. In fact, we find that the defect enables an anyon to bind to an ordinary trion for a broad range of parameters, with energy close to the trapped trion energy.  

\section{\label{sec: pl red-shift}Photoluminescence red-shift}

In order to resolve the observable PL frequency of such a bound anyonic molecule, however, we must define the final state it decays to after being optically excited. We therefore propose a minimal scenario for anyonic molecule formation. Shining light on a defect induces a bound exciton with energy $E_X$, which can lower its energy by attaching to the positive charge contained in the defect region. 
The most significant energy reduction is achieved if the defect contains a single hole -- which we assume is localized into the region of the trapping potential -- leading to a bound trion with energy $E_T \ll E_X$. If the potential well is not so deep as to repel the electron contained in the trion, the bound trion energy $E_T$ is also reduced from its ``free'' value without a trapping potential, $E_T<E^{\mathrm{free}}_{{T}}$. 
 Note that the energies above are measured with respect to the band gap, denoted $E_{bg}$. 

When the exciton recombines, a photon is emitted with frequency $\omega$, and the hole relaxes back into the defect potential, where it has energy $E_{h}$. The emitted photon energy for the bound trion is thus the difference between these two states: ${\hbar\,\omega_{T}=E_T-E_h + E_{bg}}$, where ${E_{bg}}$ is the band gap. 
The trapped trion frequency is red-shifted compared to the free trion except for very deep, narrow trapping potentials (see App.~\ref{app: AdditionalData}).
This frequency shift may be understood as the difference between the ground state energies of the two few-body problems.

A similar scenario applies to the anyonic molecule. Doping holes into the FQAH plateau populates the system with itinerant anyons. On exciting the bound trion with light, it can in principle lower its binding energy further to $E_{aT}$ by picking up one of these anyons. Because of its repulsion from the trion core, this anyon will remain delocalized on the scale of the moiré superlattice, with its wave function forming a  ``halo'' circling the trion (See App.~\ref{app: AdditionalData}). In this case, recombination of the exciton will cause both the anyon and the hole to sink back into the defect region with energy $E_{ah}$. The photon emitted in this process will therefore have energy,
\begin{align}
\hbar\,\omega_{aT}=E_{bg}+E_{aT}-E_{ah}\,,
\end{align}
which can be computed numerically by solving the corresponding few-body quantum mechanics problems. However, we emphasize that $E_{ah}$ is an estimate based on the assumption that the hole -- which originated from the FQAH liquid -- does not re-fractionalize into anyons on relaxing into the defect potential. Such a situation would in principle result in corrections due to the anyons' electrostatic repulsion and statistical interactions. 

Because the trapping potential we consider is attractive for \emph{positive} charges, we expect the anyonic molecule to have lower energy than the bound trion, $E_{aT}<E_{T}$. 
The more experimentally relevant question, however, is whether the emitted \emph{photon frequency} $\omega_{aT}$ is also lower, since it incorporates information about both the initial and final states in the vicinity of the defect. Hence, we are particularly interested in the difference between the anyonic molecule and bound trion PL frequencies,
\begin{align}\label{eq:Shift}
\hbar\,\Delta\omega_a=(E_T-E_h)-(E_{aT}-E_{ah})\,,
\end{align}
which we note is independent of the band gap.  

\begin{figure}[t]
    \centering
    \includegraphics[width=0.9\linewidth]{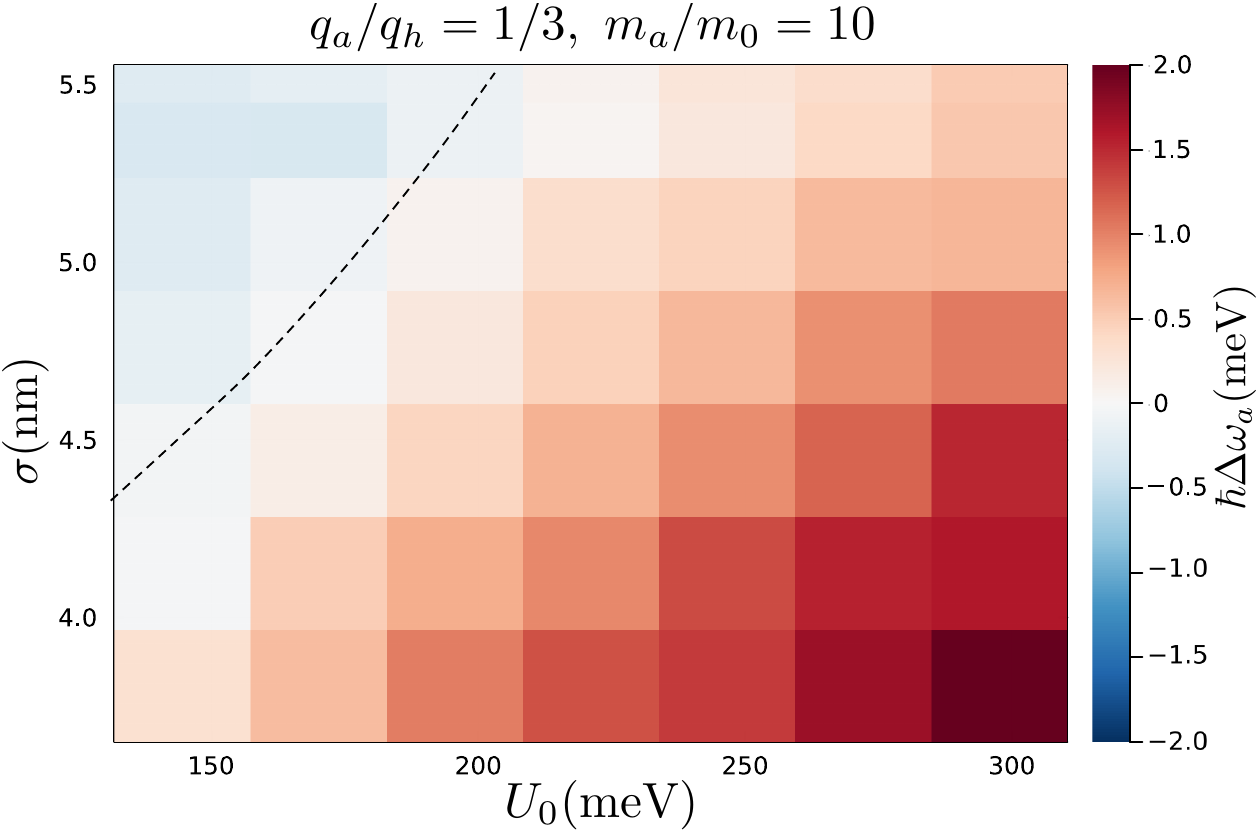}
    \caption{
    Anyon-induced shift $\hbar\Delta\omega_a$[ Eq.~\eqref{eq:Shift}] of the anyonic molecule PL peak relative to the bound trion, as a function of the bare potential depth $U_0$ and trap size $\sigma$. The dashed line marks $\hbar \Delta \omega_a = 0$, demarcating the boundary between red-shift $(\hbar\Delta\omega_a>0)$ and blue-shift $(\hbar\Delta\omega_a<0)$. Note that the bare potential $U_0$ is not to be confused with the defect energy scale felt by the anyon, which is the combined effect of the bare potential with a trapped hole or trion.
    } \label{fig:V0_vs_sigma_stability}
\end{figure}

In Fig.~\ref{fig:V0_vs_sigma_stability}, we plot $\hbar\,\Delta\omega_a$ for an anyon with charge-${e/3}$ and effective mass, $m_a = 10\, m_0$, over a range of trap depths, $U_0$ and widths $\sigma$. We find that a red-shifted ($\hbar\Delta\omega_a>0$) anyonic molecule is favored for deeper, narrower traps. On the other hand, larger, shallower traps lead to a blue-shift ($\hbar\Delta\omega_a<0$), indicating that the mutual Coulomb repulsion between the anyon and hole outweighs their attraction to the trap. 

\section{Effective defect potential}

With the parameter choices discussed above, defects of order the moir\'{e} unit cell size -- a few nanometers -- can support a red-shifted anyonic molecule. Although the needed energy scale $U_0$ in this range (see Fig.~\ref{fig:V0_vs_sigma_stability}) appears very large, the resulting red-shift is only of $\mathcal{O}(1~\mathrm{meV})$. 
This dramatic reduction can be understood by considering the electrostatic environment felt by the anyon in the region of the trap.

\begin{figure}[t]
    \centering
    \includegraphics[width=0.9\linewidth]{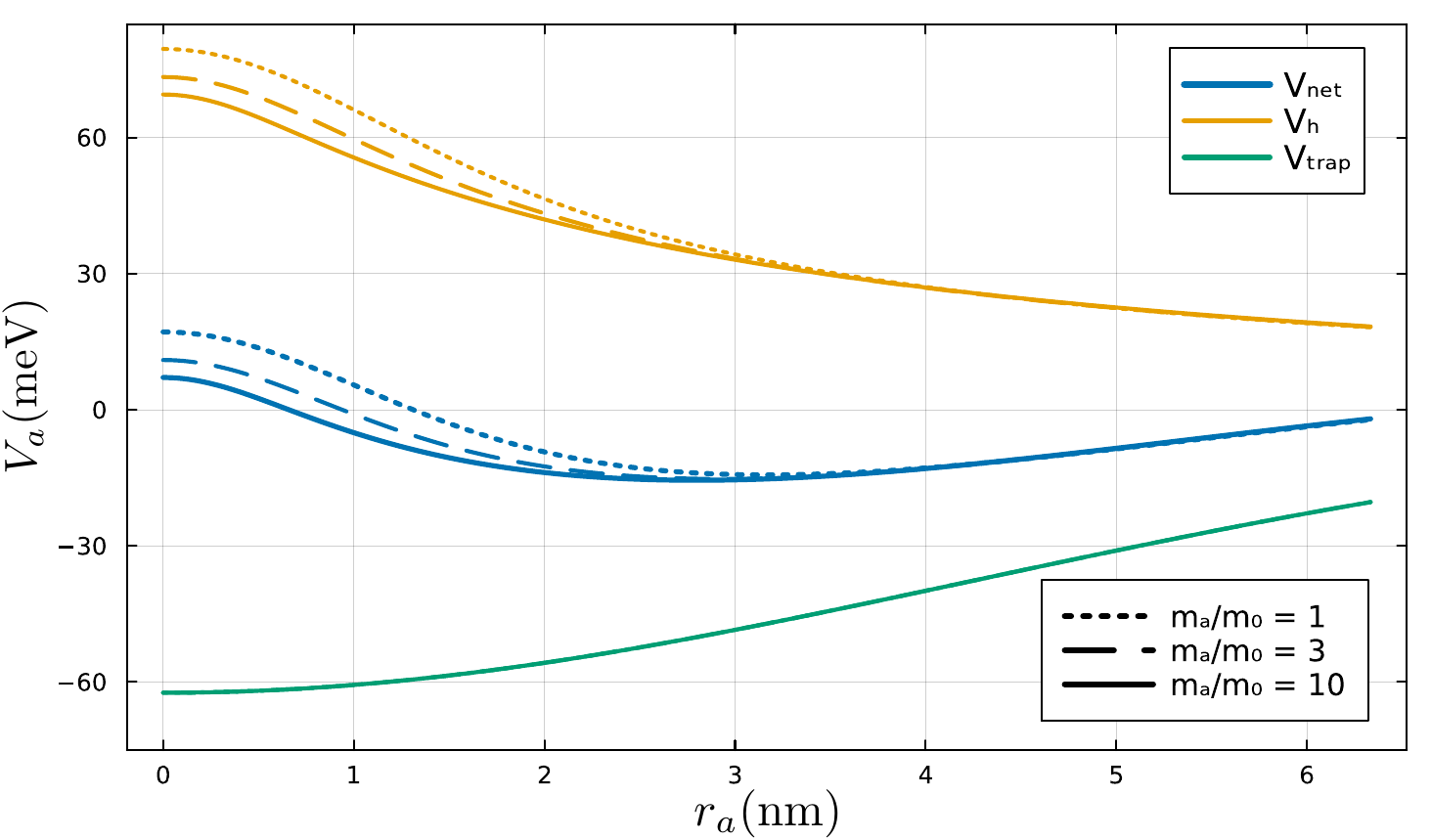}
    \includegraphics[width=0.9\linewidth]{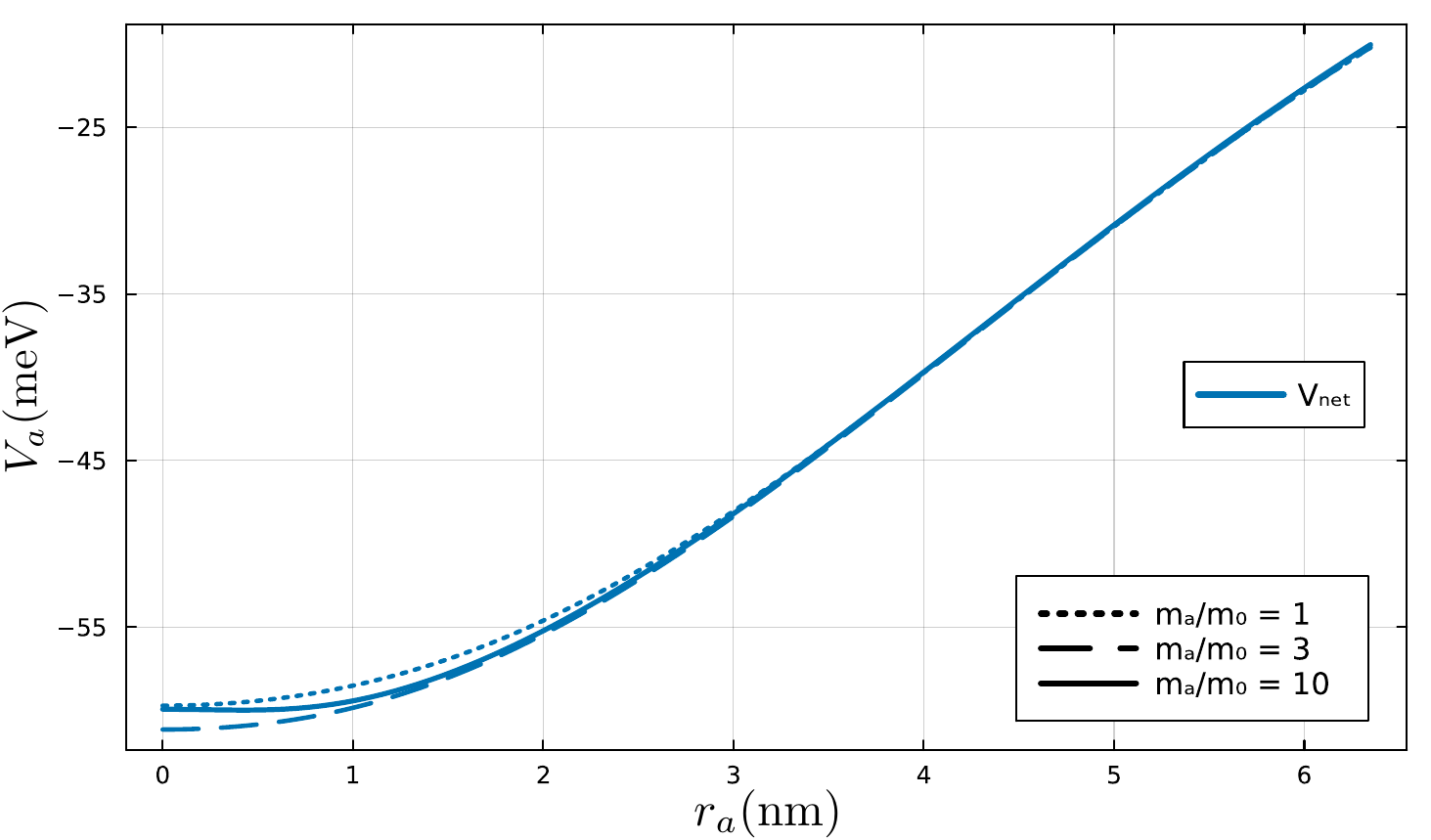}
    \caption{Effective potential on the anyon (${q_a/q_h=1/3}$) for different anyon masses in the anyon-hole bound state (top panel) and the anyon-trion bound state (bottom panel). $r_a$ is the absolute distance from the center of the trap. ${V_{\text{net}}=V_{\text{trap}}+V_{\rm{h}}}$ for the anyon-hole state, and ${V_{\text{net}}=V_{\text{trap}}+2V_{\rm{h}}+V_{\rm{e}}}$ for the anyon-trion state. Plots are evaluated with trap parameters ${U_0=190}$~meV and ${\sigma=4.2}$~nm.}
    \label{fig:Anyon Effective Potential}
\end{figure}

Indeed, the presence of a hole weakens the effect of the trap, causing the anyon to experience a screened potential of similar order of magnitude to the FQAH gap, ${\sim 5 - 10~\mathrm{meV}}$ for the ${\nu=2/3}$ state~\cite{Reddy2023,Reddy2023a}. To quantify this effect,
we define the net potential felt by the anyon due to the defect (trapping potential and hole) as
\begin{align}
    V_{\text{net}}(\rbs_a)=V_{\rm{trap}}(\rbs_a) + V_{\rm{h}}(\rbs_a), 
\end{align}
where $V_{\rm{trap}}(\rbs_a) = -q_a U(\rbs_a)$ is the potential due to the trap, and $V_{\rm{h}}$ is the Coulomb potential generated by the trapped hole:
\begin{equation}
    V_{\rm{h}}(\rbs_a) = \frac{q_aq_h\int V(\abs{\rbs_a-\rbs_h})\, 
|\psi(\rbs_a,\rbs_h)|^2\, \dd^2{\rbs_h}}{\int |\psi(\rbs_a,\rbs_h)|^2 \dd^2\rbs_h}.
\end{equation}
Here $\psi(\rbs_a,\rbs_h)$ is the wave function of the trapped anyon-hole and $V(r)$ is the Rytova-Keldysh potential (Eq.~\ref{eq:VRK:Ineraction}). The calculated potentials are shown in Fig.~\ref{fig:Anyon Effective Potential}. Owing to rotational invariance, the potential depends only on the radial coordinate $r_a \equiv \abs{\rbs_a}$. 

In Fig.~\ref{fig:Anyon Effective Potential}, it is clear that $V_{\mathrm{net}}$ has a minimum at ${r_a^{\mathrm{min}}\approx 2~\mathrm{nm}}$, where the anyon density peaks. On increasing the effective mass $m_a$, the anyon gradually localizes around the potential minimum, and its overlap with the hole is reduced. Consequently, $V_{\rm{h}}$ is reduced at short separations (See Fig.~\ref{fig:anyon density profile} of App.~\ref{app: AdditionalData} for the anyon density profile). 

This analysis applies to the problem of a trapped hole interacting with an anyon, which is the ``final'' state in our model after photon emission. To complete the discussion we must also consider the ``initial'' state, where the hole is replaced by a trapped trion. We define the net potential on the anyon due to a trion by 
\begin{align}
    V_{\text{net}}(\rbs_a)=V_{\rm{trap}}(\rbs_a) + 2V_{\rm{h}}(\rbs_a)+V_{\rm{e}}(\rbs_a), 
\end{align}
where $V_{i}$ for $i=\rm{h,e}$ is the Coulomb potential by the trapped hole/electron on the anyon,
\begin{align}
    V_{i}(\rbs_a) &= \frac{q_aq_{i}\int V(\abs{\rbs_a-\rbs_{i}})\, \rho_{ai}(\rbs_a,\rbs_i)\dd^2 \rbs_i}{\int \rho_{ai}(\rbs_a,\rbs_i)\dd^2 \rbs_i},\\
    \rho_{ai}(\rbs_a,\rbs_i)&=\int|\psi(\rbs_a,\rbs_{h1},\rbs_{h2},\rbs_e)|^2 \prod_{\beta\neq a,i}\dd^2\rbs_{\beta}.
\end{align}
Note that $V_{\rm{h}_1}=V_{\rm{h}_1}=V_{\rm{h}}$ by the symmetry between $\rm{h}_1$ and $\rm{h}_2$. We plot the resulting potential in Fig.~\ref{fig:Anyon Effective Potential}. Here the potential is only reduced by a factor of three, in contrast with the hole-anyon case above, wherein the reduction is by about an order of magnitude. Although the resulting energy scales are several times the expected FQAH gap, our model does not actually contain any information about such many-body energy scales. A more complete microscopic model incorporating the correlations of the broader FQAH fluid could predict further reductions in the anyon-trion interaction scale.

\begin{figure*}[t]
    \centering
    \includegraphics[width=0.45\linewidth]{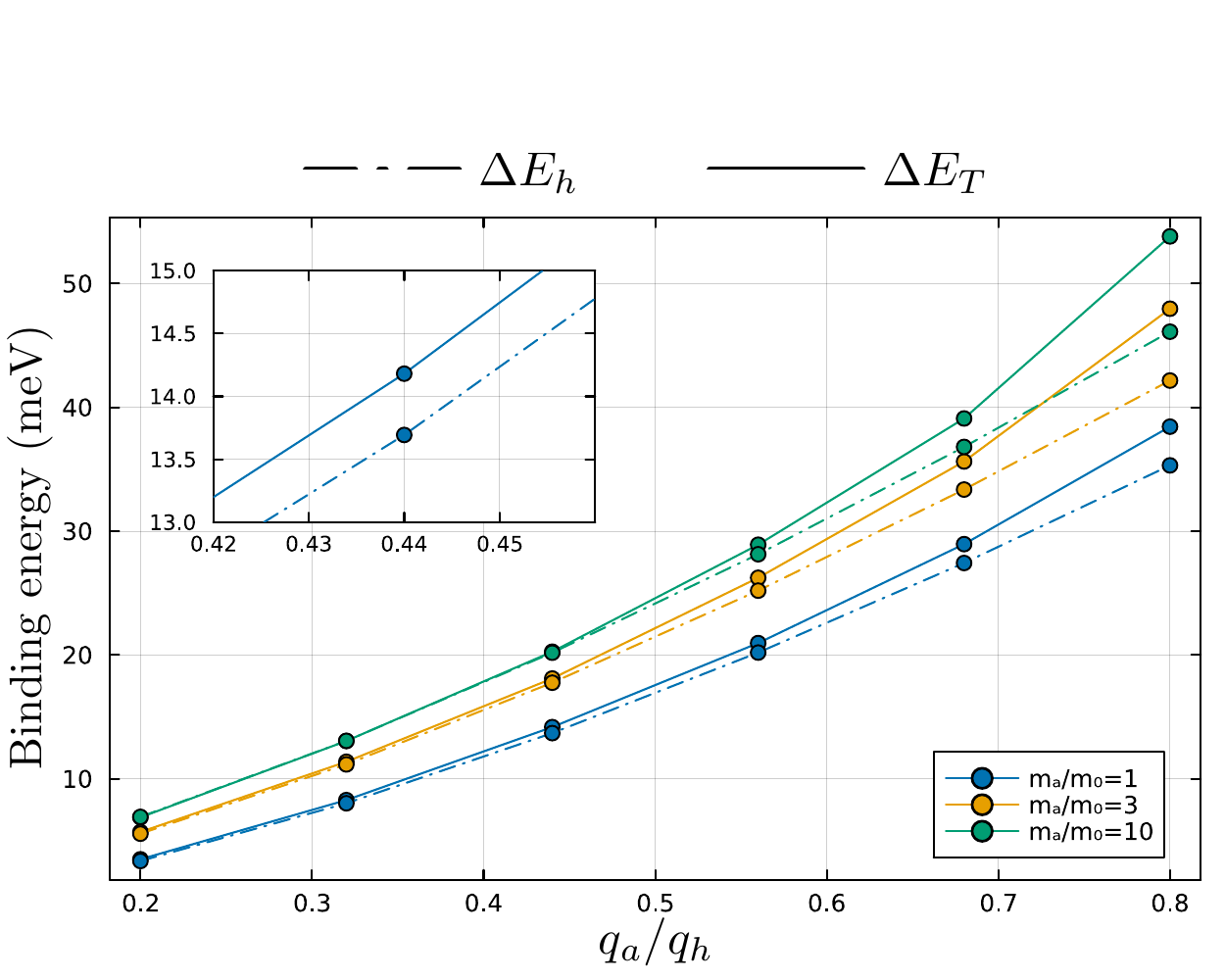}
    \includegraphics[width=0.45\linewidth]{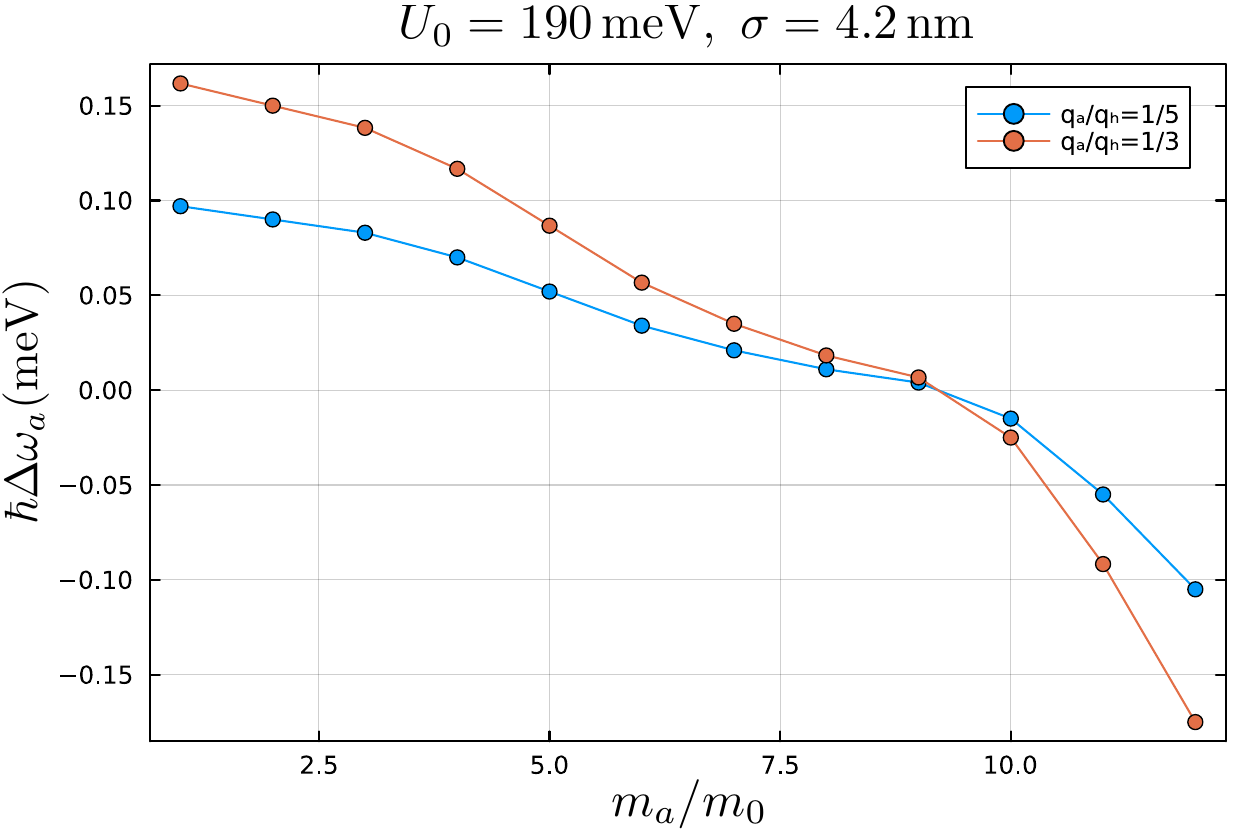}
    \caption{\label{fig:EnergyVSmass} 
    Energetics of bound states in the trap with parameters $U_0 = 190\rm{meV}$ and $\sigma=4.2\rm{nm}$. Left panel: binding energy of the anyon to the trapped hole $\Delta E_h$ and trapped trion $\Delta E_t$ (Eq.~(\ref{eq:AnyonBindingEnergies}) in the main text). The binding energies exhibit linear dependence for $q_a/q_h\lesssim 2/5$.
    Inset: guide to the eye for difference in energy between the binding energies for $m_a/m_0=1$ and $q_a/q_h\approx 4/9$.    
    Right panel: difference between emitted photon energies with and without a valence anyon (c.f. Eq.~\ref{eq:Shift} in the main text) as a function of anyon mass $m_a$ in units of the bare electron mass $m_0$, for two representative values of the anyon charge $q_a$ in units of the hole charge $q_h$. The anyonic molecule is red-shifted ($\Delta\omega_a>0$) for small $m_a/m_0$, and crosses to a blue-shifted regime at $m_a/m_0\approx 9$. }
\end{figure*}

\section{\label{sec:fractional charge}Glimpsing anyon dynamics}

Our analysis indicates that a red-shift in the trion spectrum near a defect is a strong indicator of the presence of a valence anyon. We may then consider how properties of this anyon are encoded in the PL shift. In particular, one may na\"{i}vely expect that the anyonic molecule's red-shift is linearly proportional to its fractional charge. If this is the case, then taking the ratio of frequency shifts for anyons of charge $q_a$ and $q_{a'}$ might yield an approximately quantized ratio of fractional charges, $q_a/q_{a'}$, as suggested in Ref.~\cite{li2026signatures}. However, how the red-shift is affected by anyon dispersion is not as clear \emph{a priori}.

We are therefore motivated to extract the dependence of the redshift, $\Delta\omega_a$, on the anyon charge and effective mass.
As we expect $\Delta\omega_a$ to be proportional to the anyon charge, $q_a$, it is natural to separate the red-shift into two contributions: 
\begin{equation}\label{eq:AnyonBindingEnergies}
    \begin{split}
    \Delta E_{T} &= \hbar(E_{T}-E_{aT})\,,\\
     \Delta E_{h} &= \hbar(E_{h}-E_{ah})\,.\\
    \end{split}
\end{equation}
The first contribution is the binding energy of the anyon to the trion in the trapping potential, while the second is the binding energy of the anyon to the hole after both are deposited in the trap. 
In Fig.~\ref{fig:EnergyVSmass}, we plot each as a function of $q_a$ and $m_a$ respectively, using trap parameters: ${\sigma=4.2}$~nm and ${U_0=190}$~meV. Both binding energies exhibit nearly linear behavior for $q_a\lesssim 2/5$ and anyon masses $m_a\geq m_0\approx 2m_{e,h}$.

The slopes of the two binding energies in Eq.~\eqref{eq:AnyonBindingEnergies} are also similar in this limit because the leading contribution to the potential generated by the trapped particle is determined by its net charge, which is identical for a trion and a hole. On increasing the value of $q_a$, we find clear non-linear features in $q_a$ due to the enhanced attraction of the anyon to the trapped electron, 
which causes reconfiguration of the anyonic molecule. 

We now consider the dependence of the frequency shift on the anyon effective mass, $m_a$. Fig.~\ref{fig:EnergyVSmass} shows that the shift ${\hbar \Delta \omega_a=\Delta E_T-\Delta E_h}$ decreases with $m_a$ and eventually becomes negative (blue-shift) at $m_a^{\rm{crit}}\approx 9m_0$. The particular value of $m_a^{\mathrm{crit}}$ depends on the details of the defect potential and should not be taken as a universal upper bound to anyonic molecule formation. We therefore conclude that a frequency red-shift depends on anyon dynamics through the effective mass, $m_a$. In fact, from Fig.~\ref{fig:EnergyVSmass} we see that in anyons of different charges $q_a, q_{a'}$ produce nearly quantized redshift ratios, ${\Delta\omega_a/\Delta\omega_{a'}\sim q_a/q_{a'}}$, if their effective masses match. But differences in their effective masses will lead to deviations from perfect quantization in our model. 

The physics of the red-shift can be understood qualitatively as follows. If $m_a$ is not too large, the anyon's wavefunction spatially extends to the defect center and overlaps significantly with other particles (see App.~\ref{app: AdditionalData}). In the final state, where the trap solely contains a hole and an anyon, this leads to a very strong mutual Coulomb repulsion. On the other hand, in the anyon-trion, we find that the electron in the trion can very effectively screen the repulsion between the anyon and the constituent holes. This results in weakened Coulomb repulsion in the anyon-trion compared to the ordinary trion, producing a red-shift $(\hbar \Delta \omega_a>0)$. 

Conversely, when $m_a$ is large, 
the anyon is well localized and is therefore insensitive to the details of the trion wave function. In this limit the trion can be effectively treated as a spatially separated positive charge repelling the anyon, and the attraction of the trapping potential alone is insufficient for achieving a red-shifted anyon-trion. We expect that a more microscopically accurate anyon wave function  taking into account its modulation on the moir\'{e} scale is needed to study binding in this regime.

\section{\label{sec:discussion} Discussion}

Our work demonstrates the possibility of anyonic molecule formation near defects in TMD FQAH systems. For our Gaussian defect model with the $t$MoTe$_2$-relevant parameters, we found the red-shift in PL peak appears when the defect width is around the $t$MoTe$_2$ moiré scale -- a few nanometers -- and the effective depth (in the presence of a trapped hole) is $\mathcal{O}(10\mathrm{meV})$, which is of similar magnitude to expectations for the $\nu=2/3$ FQAH gap. This general consistency is perhaps surprising, given that our phenomenological model lacks detailed microscopic input and relies on a number of assumptions, such as the particular initial and final states and the  form for the defect potential. Indeed, the nature of defects in TMD systems is a subject still under experimental investigation.

Hence the lesson of our minimal few-body model is remarkably robust. Namely, that defects provide a natural optical laboratory for itinerant anyons in moiré FQAH materials, providing a promising route toward anyon control. Indeed, one may even consider patterning FQAH systems with nano-scale anyon traps in the laboratory by bringing an atomic-force microscope tip in contact with a FQAH sample, forming correlated arrays of anyon traps atop the FQAH background. The physics and measurable properties of such systems has not been extensively explored, offering fertile ground for future exploration.

Our model is consistent with the basic experimental conclusions of Ref.~\cite{li2026signatures}, where it was initially speculated that a PL red-shift can signal the binding of fractionally quantized charges to defects. However, we do not find evidence of a universal red-shift ratio between plateaus, at least for the parameter regime with a red-shifted anyon-trion bound state. Instead, we find a dependence on the anyon dynamics in our model: The red-shift depends significantly on the anyon effective mass, as can be seen from inspecting Fig.~\ref{fig:EnergyVSmass}. This means that differences in the anyon dispersion between plateaus can lead to an unquantized red-shift ratio.

Our methods also present opportunities to study the optical signatures of multi-anyon molecules, where braiding and fusion effects become essential. The formation of such composites is also a real possibility, and determining how to distinguish them from the simpler anyon-trions in this work is an urgent and important challenge.

\begin{acknowledgments}
We are especially grateful to Xiao-Dong Xu for inspiring discussions and to Daniel Parker and Alex Thomson for comments on the manuscript. We also thank Valentin Cr\'{e}pel, Steve Kivelson, Dahlia Klein, Weijie Li, Sri Raghu, Zengde She, Boris Shklovskii, and Yue Zhao for discussions.
This work was supported by startup funds at the University of Minnesota (HRC, UM, and HG) and by the Bill Fine Postdoctoral Fellowship (VC). 

\emph{Note added.} After this work was completed, we became aware of complementary work in Ref.~\cite{Wagner2026}, which also considers the problem of anyons binding to defects of like charge using different methods. 
\end{acknowledgments}

\bibliography{references}

\begin{thebibliography}{52}%
\makeatletter
\providecommand \@ifxundefined [1]{%
 \@ifx{#1\undefined}
}%
\providecommand \@ifnum [1]{%
 \ifnum #1\expandafter \@firstoftwo
 \else \expandafter \@secondoftwo
 \fi
}%
\providecommand \@ifx [1]{%
 \ifx #1\expandafter \@firstoftwo
 \else \expandafter \@secondoftwo
 \fi
}%
\providecommand \natexlab [1]{#1}%
\providecommand \enquote  [1]{``#1''}%
\providecommand \bibnamefont  [1]{#1}%
\providecommand \bibfnamefont [1]{#1}%
\providecommand \citenamefont [1]{#1}%
\providecommand \href@noop [0]{\@secondoftwo}%
\providecommand \href [0]{\begingroup \@sanitize@url \@href}%
\providecommand \@href[1]{\@@startlink{#1}\@@href}%
\providecommand \@@href[1]{\endgroup#1\@@endlink}%
\providecommand \@sanitize@url [0]{\catcode `\\12\catcode `\$12\catcode `\&12\catcode `\#12\catcode `\^12\catcode `\_12\catcode `\%12\relax}%
\providecommand \@@startlink[1]{}%
\providecommand \@@endlink[0]{}%
\providecommand \url  [0]{\begingroup\@sanitize@url \@url }%
\providecommand \@url [1]{\endgroup\@href {#1}{\urlprefix }}%
\providecommand \urlprefix  [0]{URL }%
\providecommand \Eprint [0]{\href }%
\providecommand \doibase [0]{https://doi.org/}%
\providecommand \selectlanguage [0]{\@gobble}%
\providecommand \bibinfo  [0]{\@secondoftwo}%
\providecommand \bibfield  [0]{\@secondoftwo}%
\providecommand \translation [1]{[#1]}%
\providecommand \BibitemOpen [0]{}%
\providecommand \bibitemStop [0]{}%
\providecommand \bibitemNoStop [0]{.\EOS\space}%
\providecommand \EOS [0]{\spacefactor3000\relax}%
\providecommand \BibitemShut  [1]{\csname bibitem#1\endcsname}%
\let\auto@bib@innerbib\@empty
\bibitem [{\citenamefont {Cai}\ \emph {et~al.}(2023)\citenamefont {Cai}, \citenamefont {Anderson}, \citenamefont {Wang}, \citenamefont {Zhang}, \citenamefont {Liu}, \citenamefont {Holtzmann}, \citenamefont {Zhang}, \citenamefont {Fan}, \citenamefont {Taniguchi}, \citenamefont {Watanabe}, \citenamefont {Ran}, \citenamefont {Cao}, \citenamefont {Fu}, \citenamefont {Xiao}, \citenamefont {Yao},\ and\ \citenamefont {Xu}}]{cai2023signatures}%
  \BibitemOpen
  \bibfield  {author} {\bibinfo {author} {\bibfnamefont {J.}~\bibnamefont {Cai}}, \bibinfo {author} {\bibfnamefont {E.}~\bibnamefont {Anderson}}, \bibinfo {author} {\bibfnamefont {C.}~\bibnamefont {Wang}}, \bibinfo {author} {\bibfnamefont {X.}~\bibnamefont {Zhang}}, \bibinfo {author} {\bibfnamefont {X.}~\bibnamefont {Liu}}, \bibinfo {author} {\bibfnamefont {W.}~\bibnamefont {Holtzmann}}, \bibinfo {author} {\bibfnamefont {Y.}~\bibnamefont {Zhang}}, \bibinfo {author} {\bibfnamefont {F.}~\bibnamefont {Fan}}, \bibinfo {author} {\bibfnamefont {T.}~\bibnamefont {Taniguchi}}, \bibinfo {author} {\bibfnamefont {K.}~\bibnamefont {Watanabe}}, \bibinfo {author} {\bibfnamefont {Y.}~\bibnamefont {Ran}}, \bibinfo {author} {\bibfnamefont {T.}~\bibnamefont {Cao}}, \bibinfo {author} {\bibfnamefont {L.}~\bibnamefont {Fu}}, \bibinfo {author} {\bibfnamefont {D.}~\bibnamefont {Xiao}}, \bibinfo {author} {\bibfnamefont {W.}~\bibnamefont {Yao}},\ and\ \bibinfo {author} {\bibfnamefont {X.}~\bibnamefont {Xu}},\ }\bibfield  {title}
  {\bibinfo {title} {Signatures of fractional quantum anomalous hall states in twisted {MoTe$_2$}},\ }\href {https://doi.org/10.1038/s41586-023-06289-w} {\bibfield  {journal} {\bibinfo  {journal} {Nature}\ }\textbf {\bibinfo {volume} {622}},\ \bibinfo {pages} {63} (\bibinfo {year} {2023})}\BibitemShut {NoStop}%
\bibitem [{\citenamefont {Zeng}\ \emph {et~al.}(2023)\citenamefont {Zeng}, \citenamefont {Xia}, \citenamefont {Kang}, \citenamefont {Zhu}, \citenamefont {Knuppel}, \citenamefont {Vaswani}, \citenamefont {Watanabe}, \citenamefont {Taniguchi}, \citenamefont {Mak},\ and\ \citenamefont {Shan}}]{zeng2023thermodynamic}%
  \BibitemOpen
  \bibfield  {author} {\bibinfo {author} {\bibfnamefont {Y.}~\bibnamefont {Zeng}}, \bibinfo {author} {\bibfnamefont {Z.}~\bibnamefont {Xia}}, \bibinfo {author} {\bibfnamefont {K.}~\bibnamefont {Kang}}, \bibinfo {author} {\bibfnamefont {J.}~\bibnamefont {Zhu}}, \bibinfo {author} {\bibfnamefont {P.}~\bibnamefont {Knuppel}}, \bibinfo {author} {\bibfnamefont {C.}~\bibnamefont {Vaswani}}, \bibinfo {author} {\bibfnamefont {K.}~\bibnamefont {Watanabe}}, \bibinfo {author} {\bibfnamefont {T.}~\bibnamefont {Taniguchi}}, \bibinfo {author} {\bibfnamefont {K.~F.}\ \bibnamefont {Mak}},\ and\ \bibinfo {author} {\bibfnamefont {J.}~\bibnamefont {Shan}},\ }\bibfield  {title} {\bibinfo {title} {Thermodynamic evidence of fractional chern insulator in moir{\'e} {MoTe$_2$}},\ }\href {https://doi.org/10.1038/s41586-023-06452-3} {\bibfield  {journal} {\bibinfo  {journal} {Nature}\ }\textbf {\bibinfo {volume} {622}},\ \bibinfo {pages} {69} (\bibinfo {year} {2023})}\BibitemShut {NoStop}%
\bibitem [{\citenamefont {Xu}\ \emph {et~al.}(2023)\citenamefont {Xu}, \citenamefont {Sun}, \citenamefont {Jia}, \citenamefont {Liu}, \citenamefont {Xu}, \citenamefont {Li}, \citenamefont {Gu}, \citenamefont {Watanabe}, \citenamefont {Taniguchi}, \citenamefont {Tong}, \citenamefont {Jia}, \citenamefont {Shi}, \citenamefont {Jiang}, \citenamefont {Zhang}, \citenamefont {Liu},\ and\ \citenamefont {Li}}]{xu2023observation}%
  \BibitemOpen
  \bibfield  {author} {\bibinfo {author} {\bibfnamefont {F.}~\bibnamefont {Xu}}, \bibinfo {author} {\bibfnamefont {Z.}~\bibnamefont {Sun}}, \bibinfo {author} {\bibfnamefont {T.}~\bibnamefont {Jia}}, \bibinfo {author} {\bibfnamefont {C.}~\bibnamefont {Liu}}, \bibinfo {author} {\bibfnamefont {C.}~\bibnamefont {Xu}}, \bibinfo {author} {\bibfnamefont {C.}~\bibnamefont {Li}}, \bibinfo {author} {\bibfnamefont {Y.}~\bibnamefont {Gu}}, \bibinfo {author} {\bibfnamefont {K.}~\bibnamefont {Watanabe}}, \bibinfo {author} {\bibfnamefont {T.}~\bibnamefont {Taniguchi}}, \bibinfo {author} {\bibfnamefont {B.}~\bibnamefont {Tong}}, \bibinfo {author} {\bibfnamefont {J.}~\bibnamefont {Jia}}, \bibinfo {author} {\bibfnamefont {Z.}~\bibnamefont {Shi}}, \bibinfo {author} {\bibfnamefont {S.}~\bibnamefont {Jiang}}, \bibinfo {author} {\bibfnamefont {Y.}~\bibnamefont {Zhang}}, \bibinfo {author} {\bibfnamefont {X.}~\bibnamefont {Liu}},\ and\ \bibinfo {author} {\bibfnamefont {T.}~\bibnamefont {Li}},\ }\bibfield  {title} {\bibinfo {title}
  {Observation of integer and fractional quantum anomalous hall effects in twisted bilayer ${\mathrm{mote}}_{2}$},\ }\href {https://doi.org/10.1103/PhysRevX.13.031037} {\bibfield  {journal} {\bibinfo  {journal} {Phys. Rev. X}\ }\textbf {\bibinfo {volume} {13}},\ \bibinfo {pages} {031037} (\bibinfo {year} {2023})}\BibitemShut {NoStop}%
\bibitem [{\citenamefont {Park}\ \emph {et~al.}(2023)\citenamefont {Park}, \citenamefont {Cai}, \citenamefont {Anderson}, \citenamefont {Zhang}, \citenamefont {Zhu}, \citenamefont {Liu}, \citenamefont {Wang}, \citenamefont {Holtzmann}, \citenamefont {Hu}, \citenamefont {Liu}, \citenamefont {Taniguchi}, \citenamefont {Watanabe}, \citenamefont {Chu}, \citenamefont {Cao}, \citenamefont {Fu}, \citenamefont {Yao}, \citenamefont {Chang}, \citenamefont {Cobden}, \citenamefont {Xiao},\ and\ \citenamefont {Xu}}]{park2023observation}%
  \BibitemOpen
  \bibfield  {author} {\bibinfo {author} {\bibfnamefont {H.}~\bibnamefont {Park}}, \bibinfo {author} {\bibfnamefont {J.}~\bibnamefont {Cai}}, \bibinfo {author} {\bibfnamefont {E.}~\bibnamefont {Anderson}}, \bibinfo {author} {\bibfnamefont {Y.}~\bibnamefont {Zhang}}, \bibinfo {author} {\bibfnamefont {J.}~\bibnamefont {Zhu}}, \bibinfo {author} {\bibfnamefont {X.}~\bibnamefont {Liu}}, \bibinfo {author} {\bibfnamefont {C.}~\bibnamefont {Wang}}, \bibinfo {author} {\bibfnamefont {W.}~\bibnamefont {Holtzmann}}, \bibinfo {author} {\bibfnamefont {C.}~\bibnamefont {Hu}}, \bibinfo {author} {\bibfnamefont {Z.}~\bibnamefont {Liu}}, \bibinfo {author} {\bibfnamefont {T.}~\bibnamefont {Taniguchi}}, \bibinfo {author} {\bibfnamefont {K.}~\bibnamefont {Watanabe}}, \bibinfo {author} {\bibfnamefont {J.-H.}\ \bibnamefont {Chu}}, \bibinfo {author} {\bibfnamefont {T.}~\bibnamefont {Cao}}, \bibinfo {author} {\bibfnamefont {L.}~\bibnamefont {Fu}}, \bibinfo {author} {\bibfnamefont {W.}~\bibnamefont {Yao}}, \bibinfo {author}
  {\bibfnamefont {C.-Z.}\ \bibnamefont {Chang}}, \bibinfo {author} {\bibfnamefont {D.}~\bibnamefont {Cobden}}, \bibinfo {author} {\bibfnamefont {D.}~\bibnamefont {Xiao}},\ and\ \bibinfo {author} {\bibfnamefont {X.}~\bibnamefont {Xu}},\ }\bibfield  {title} {\bibinfo {title} {Observation of fractionally quantized anomalous hall effect},\ }\href {https://doi.org/10.1038/s41586-023-06536-0} {\bibfield  {journal} {\bibinfo  {journal} {Nature}\ }\textbf {\bibinfo {volume} {622}},\ \bibinfo {pages} {74} (\bibinfo {year} {2023})}\BibitemShut {NoStop}%
\bibitem [{\citenamefont {Xu}\ \emph {et~al.}(2025{\natexlab{a}})\citenamefont {Xu}, \citenamefont {Sun}, \citenamefont {Li}, \citenamefont {Zheng}, \citenamefont {Xu}, \citenamefont {Gao}, \citenamefont {Jia}, \citenamefont {Watanabe}, \citenamefont {Taniguchi}, \citenamefont {Tong}, \citenamefont {Lu}, \citenamefont {Jia}, \citenamefont {Shi}, \citenamefont {Jiang}, \citenamefont {Zhang}, \citenamefont {Zhang}, \citenamefont {Lei}, \citenamefont {Liu},\ and\ \citenamefont {Li}}]{Li2025}%
  \BibitemOpen
  \bibfield  {author} {\bibinfo {author} {\bibfnamefont {F.}~\bibnamefont {Xu}}, \bibinfo {author} {\bibfnamefont {Z.}~\bibnamefont {Sun}}, \bibinfo {author} {\bibfnamefont {J.}~\bibnamefont {Li}}, \bibinfo {author} {\bibfnamefont {C.}~\bibnamefont {Zheng}}, \bibinfo {author} {\bibfnamefont {C.}~\bibnamefont {Xu}}, \bibinfo {author} {\bibfnamefont {J.}~\bibnamefont {Gao}}, \bibinfo {author} {\bibfnamefont {T.}~\bibnamefont {Jia}}, \bibinfo {author} {\bibfnamefont {K.}~\bibnamefont {Watanabe}}, \bibinfo {author} {\bibfnamefont {T.}~\bibnamefont {Taniguchi}}, \bibinfo {author} {\bibfnamefont {B.}~\bibnamefont {Tong}}, \bibinfo {author} {\bibfnamefont {L.}~\bibnamefont {Lu}}, \bibinfo {author} {\bibfnamefont {J.}~\bibnamefont {Jia}}, \bibinfo {author} {\bibfnamefont {Z.}~\bibnamefont {Shi}}, \bibinfo {author} {\bibfnamefont {S.}~\bibnamefont {Jiang}}, \bibinfo {author} {\bibfnamefont {Y.}~\bibnamefont {Zhang}}, \bibinfo {author} {\bibfnamefont {Y.}~\bibnamefont {Zhang}}, \bibinfo {author} {\bibfnamefont
  {S.}~\bibnamefont {Lei}}, \bibinfo {author} {\bibfnamefont {X.}~\bibnamefont {Liu}},\ and\ \bibinfo {author} {\bibfnamefont {T.}~\bibnamefont {Li}},\ }\bibfield  {title} {\bibinfo {title} {Signatures of unconventional superconductivity near reentrant and fractional quantum anomalous {Hall} insulators},\ }\href {https://arxiv.org/abs/2504.06972} {\bibfield  {journal} {\bibinfo  {journal} {arXiv preprint arXiv:2504.06972}\ } (\bibinfo {year} {2025}{\natexlab{a}})}\BibitemShut {NoStop}%
\bibitem [{\citenamefont {Tang}\ and\ \citenamefont {Wen}(2013)}]{tang2013superconductivity}%
  \BibitemOpen
  \bibfield  {author} {\bibinfo {author} {\bibfnamefont {E.}~\bibnamefont {Tang}}\ and\ \bibinfo {author} {\bibfnamefont {X.-G.}\ \bibnamefont {Wen}},\ }\bibfield  {title} {\bibinfo {title} {Superconductivity with intrinsic topological order induced by pure coulomb interaction and time-reversal symmetry breaking},\ }\href {https://doi.org/10.1103/PhysRevB.88.195117} {\bibfield  {journal} {\bibinfo  {journal} {Phys. Rev. B}\ }\textbf {\bibinfo {volume} {88}},\ \bibinfo {pages} {195117} (\bibinfo {year} {2013})}\BibitemShut {NoStop}%
\bibitem [{\citenamefont {Kim}\ \emph {et~al.}(2025)\citenamefont {Kim}, \citenamefont {Timmel}, \citenamefont {Ju},\ and\ \citenamefont {Wen}}]{kim2025topological}%
  \BibitemOpen
  \bibfield  {author} {\bibinfo {author} {\bibfnamefont {M.}~\bibnamefont {Kim}}, \bibinfo {author} {\bibfnamefont {A.}~\bibnamefont {Timmel}}, \bibinfo {author} {\bibfnamefont {L.}~\bibnamefont {Ju}},\ and\ \bibinfo {author} {\bibfnamefont {X.-G.}\ \bibnamefont {Wen}},\ }\bibfield  {title} {\bibinfo {title} {Topological chiral superconductivity beyond pairing in a fermi liquid},\ }\href {https://doi.org/10.1103/PhysRevB.111.014508} {\bibfield  {journal} {\bibinfo  {journal} {Phys. Rev. B}\ }\textbf {\bibinfo {volume} {111}},\ \bibinfo {pages} {014508} (\bibinfo {year} {2025})}\BibitemShut {NoStop}%
\bibitem [{\citenamefont {Schleith}\ \emph {et~al.}(2025)\citenamefont {Schleith}, \citenamefont {Soejima},\ and\ \citenamefont {Khalaf}}]{schleith2025anyon}%
  \BibitemOpen
  \bibfield  {author} {\bibinfo {author} {\bibfnamefont {M.-L.}\ \bibnamefont {Schleith}}, \bibinfo {author} {\bibfnamefont {T.}~\bibnamefont {Soejima}},\ and\ \bibinfo {author} {\bibfnamefont {E.}~\bibnamefont {Khalaf}},\ }\bibfield  {title} {\bibinfo {title} {Anyon dispersion from non-uniform magnetic field on the sphere},\ }\href {https://arxiv.org/abs/2506.11211} {\bibfield  {journal} {\bibinfo  {journal} {arXiv preprint arXiv:2506.11211}\ } (\bibinfo {year} {2025})}\BibitemShut {NoStop}%
\bibitem [{\citenamefont {Yan}\ \emph {et~al.}(2025)\citenamefont {Yan}, \citenamefont {Li}, \citenamefont {Soejima},\ and\ \citenamefont {Khalaf}}]{yan2025anyon}%
  \BibitemOpen
  \bibfield  {author} {\bibinfo {author} {\bibfnamefont {Z.}~\bibnamefont {Yan}}, \bibinfo {author} {\bibfnamefont {Q.}~\bibnamefont {Li}}, \bibinfo {author} {\bibfnamefont {T.}~\bibnamefont {Soejima}},\ and\ \bibinfo {author} {\bibfnamefont {E.}~\bibnamefont {Khalaf}},\ }\bibfield  {title} {\bibinfo {title} {Anyon dispersion in {Aharonov-Casher} bands and implications for twisted mote2},\ }\href {https://arxiv.org/abs/2512.15863} {\bibfield  {journal} {\bibinfo  {journal} {arXiv preprint arXiv:2512.15863}\ } (\bibinfo {year} {2025})}\BibitemShut {NoStop}%
\bibitem [{\citenamefont {Shi}\ and\ \citenamefont {Senthil}(2025)}]{shi2025doping}%
  \BibitemOpen
  \bibfield  {author} {\bibinfo {author} {\bibfnamefont {Z.~D.}\ \bibnamefont {Shi}}\ and\ \bibinfo {author} {\bibfnamefont {T.}~\bibnamefont {Senthil}},\ }\bibfield  {title} {\bibinfo {title} {Doping a fractional quantum anomalous hall insulator},\ }\href {https://journals.aps.org/prx/abstract/10.1103/kcm5-hx56} {\bibfield  {journal} {\bibinfo  {journal} {Phys. Rev. X}\ }\textbf {\bibinfo {volume} {15}},\ \bibinfo {pages} {031069} (\bibinfo {year} {2025})}\BibitemShut {NoStop}%
\bibitem [{\citenamefont {Divic}\ \emph {et~al.}(2025)\citenamefont {Divic}, \citenamefont {Cr{\'e}pel}, \citenamefont {Soejima}, \citenamefont {Song}, \citenamefont {Millis}, \citenamefont {Zaletel},\ and\ \citenamefont {Vishwanath}}]{divic2025anyon}%
  \BibitemOpen
  \bibfield  {author} {\bibinfo {author} {\bibfnamefont {S.}~\bibnamefont {Divic}}, \bibinfo {author} {\bibfnamefont {V.}~\bibnamefont {Cr{\'e}pel}}, \bibinfo {author} {\bibfnamefont {T.}~\bibnamefont {Soejima}}, \bibinfo {author} {\bibfnamefont {X.-Y.}\ \bibnamefont {Song}}, \bibinfo {author} {\bibfnamefont {A.~J.}\ \bibnamefont {Millis}}, \bibinfo {author} {\bibfnamefont {M.~P.}\ \bibnamefont {Zaletel}},\ and\ \bibinfo {author} {\bibfnamefont {A.}~\bibnamefont {Vishwanath}},\ }\bibfield  {title} {\bibinfo {title} {Anyon superconductivity from topological criticality in a hofstadter--hubbard model},\ }\href {https://www.pnas.org/doi/10.1073/pnas.2426680122} {\bibfield  {journal} {\bibinfo  {journal} {Proceedings of the National Academy of Sciences}\ }\textbf {\bibinfo {volume} {122}},\ \bibinfo {pages} {e2426680122} (\bibinfo {year} {2025})}\BibitemShut {NoStop}%
\bibitem [{\citenamefont {Shi}\ \emph {et~al.}(2025)\citenamefont {Shi}, \citenamefont {Zhang},\ and\ \citenamefont {Senthil}}]{shi2025dopingB}%
  \BibitemOpen
  \bibfield  {author} {\bibinfo {author} {\bibfnamefont {Z.~D.}\ \bibnamefont {Shi}}, \bibinfo {author} {\bibfnamefont {C.}~\bibnamefont {Zhang}},\ and\ \bibinfo {author} {\bibfnamefont {T.}~\bibnamefont {Senthil}},\ }\bibfield  {title} {\bibinfo {title} {Doping lattice non-abelian quantum hall states},\ }\href {https://scipost.org/SciPostPhys.19.6.150} {\bibfield  {journal} {\bibinfo  {journal} {SciPost Physics}\ }\textbf {\bibinfo {volume} {19}},\ \bibinfo {pages} {150} (\bibinfo {year} {2025})}\BibitemShut {NoStop}%
\bibitem [{\citenamefont {Pichler}\ \emph {et~al.}(2025)\citenamefont {Pichler}, \citenamefont {Kuhlenkamp}, \citenamefont {Knap},\ and\ \citenamefont {Vishwanath}}]{pichler2025microscopic}%
  \BibitemOpen
  \bibfield  {author} {\bibinfo {author} {\bibfnamefont {F.}~\bibnamefont {Pichler}}, \bibinfo {author} {\bibfnamefont {C.}~\bibnamefont {Kuhlenkamp}}, \bibinfo {author} {\bibfnamefont {M.}~\bibnamefont {Knap}},\ and\ \bibinfo {author} {\bibfnamefont {A.}~\bibnamefont {Vishwanath}},\ }\bibfield  {title} {\bibinfo {title} {Microscopic mechanism of anyon superconductivity emerging from fractional chern insulators},\ }\href {https://www.cell.com/newton/fulltext/S2950-6360(25)00332-9} {\bibfield  {journal} {\bibinfo  {journal} {Newton}\ } (\bibinfo {year} {2025})}\BibitemShut {NoStop}%
\bibitem [{\citenamefont {Gon{\c{c}}alves}\ \emph {et~al.}(2025)\citenamefont {Gon{\c{c}}alves}, \citenamefont {Mendez-Valderrama}, \citenamefont {Herzog-Arbeitman}, \citenamefont {Yu}, \citenamefont {Xu}, \citenamefont {Xiao}, \citenamefont {Bernevig},\ and\ \citenamefont {Regnault}}]{gonccalves2025spinless}%
  \BibitemOpen
  \bibfield  {author} {\bibinfo {author} {\bibfnamefont {M.}~\bibnamefont {Gon{\c{c}}alves}}, \bibinfo {author} {\bibfnamefont {J.~F.}\ \bibnamefont {Mendez-Valderrama}}, \bibinfo {author} {\bibfnamefont {J.}~\bibnamefont {Herzog-Arbeitman}}, \bibinfo {author} {\bibfnamefont {J.}~\bibnamefont {Yu}}, \bibinfo {author} {\bibfnamefont {X.}~\bibnamefont {Xu}}, \bibinfo {author} {\bibfnamefont {D.}~\bibnamefont {Xiao}}, \bibinfo {author} {\bibfnamefont {B.~A.}\ \bibnamefont {Bernevig}},\ and\ \bibinfo {author} {\bibfnamefont {N.}~\bibnamefont {Regnault}},\ }\bibfield  {title} {\bibinfo {title} {Spinless and spinful charge excitations in moir{\'e} fractional chern insulators},\ }\href {https://arxiv.org/abs/2506.05330} {\bibfield  {journal} {\bibinfo  {journal} {arXiv preprint arXiv:2506.05330}\ } (\bibinfo {year} {2025})}\BibitemShut {NoStop}%
\bibitem [{\citenamefont {Iyer}\ \emph {et~al.}(2026)\citenamefont {Iyer}, \citenamefont {Feuerpfeil}, \citenamefont {Crépel}, \citenamefont {Regnault},\ and\ \citenamefont {Mora}}]{Iyer2026}%
  \BibitemOpen
  \bibfield  {author} {\bibinfo {author} {\bibfnamefont {K.}~\bibnamefont {Iyer}}, \bibinfo {author} {\bibfnamefont {A.}~\bibnamefont {Feuerpfeil}}, \bibinfo {author} {\bibfnamefont {V.}~\bibnamefont {Crépel}}, \bibinfo {author} {\bibfnamefont {N.}~\bibnamefont {Regnault}},\ and\ \bibinfo {author} {\bibfnamefont {C.}~\bibnamefont {Mora}},\ }\href {https://arxiv.org/abs/2604.24859} {\bibinfo {title} {Dispersion of anyon bloch bands}} (\bibinfo {year} {2026}),\ \Eprint {https://arxiv.org/abs/2604.24859} {arXiv:2604.24859 [cond-mat.mes-hall]} \BibitemShut {NoStop}%
\bibitem [{\citenamefont {Rashba}\ and\ \citenamefont {Portnoi}(1993)}]{rashba1993anyon}%
  \BibitemOpen
  \bibfield  {author} {\bibinfo {author} {\bibfnamefont {E.~I.}\ \bibnamefont {Rashba}}\ and\ \bibinfo {author} {\bibfnamefont {M.~E.}\ \bibnamefont {Portnoi}},\ }\bibfield  {title} {\bibinfo {title} {Anyon excitons},\ }\href@noop {} {\bibfield  {journal} {\bibinfo  {journal} {Phys. Rev. Lett.}\ }\textbf {\bibinfo {volume} {70}},\ \bibinfo {pages} {3315} (\bibinfo {year} {1993})}\BibitemShut {NoStop}%
\bibitem [{\citenamefont {W\'ojs}\ and\ \citenamefont {Quinn}(2000)}]{Wojs2000}%
  \BibitemOpen
  \bibfield  {author} {\bibinfo {author} {\bibfnamefont {A.}~\bibnamefont {W\'ojs}}\ and\ \bibinfo {author} {\bibfnamefont {J.~J.}\ \bibnamefont {Quinn}},\ }\bibfield  {title} {\bibinfo {title} {Energy spectra of fractional quantum hall systems in the presence of a valence hole},\ }\href {https://doi.org/10.1103/PhysRevB.63.045303} {\bibfield  {journal} {\bibinfo  {journal} {Phys. Rev. B}\ }\textbf {\bibinfo {volume} {63}},\ \bibinfo {pages} {045303} (\bibinfo {year} {2000})}\BibitemShut {NoStop}%
\bibitem [{\citenamefont {Wagner}\ and\ \citenamefont {Neupert}(2026)}]{Wagner:2025kig}%
  \BibitemOpen
  \bibfield  {author} {\bibinfo {author} {\bibfnamefont {G.}~\bibnamefont {Wagner}}\ and\ \bibinfo {author} {\bibfnamefont {T.}~\bibnamefont {Neupert}},\ }\bibfield  {title} {\bibinfo {title} {{Sensing the binding and unbinding of anyons at impurities}},\ }\href {https://doi.org/10.1103/cfm5-wgz1} {\bibfield  {journal} {\bibinfo  {journal} {Phys. Rev. Res.}\ }\textbf {\bibinfo {volume} {8}},\ \bibinfo {pages} {013263} (\bibinfo {year} {2026})},\ \Eprint {https://arxiv.org/abs/2507.08928} {arXiv:2507.08928 [cond-mat.str-el]} \BibitemShut {NoStop}%
\bibitem [{\citenamefont {Paul}\ \emph {et~al.}(2025)\citenamefont {Paul}, \citenamefont {Abouelkomsan}, \citenamefont {Reddy},\ and\ \citenamefont {Fu}}]{paul2025shininglightcollectivemodes}%
  \BibitemOpen
  \bibfield  {author} {\bibinfo {author} {\bibfnamefont {N.}~\bibnamefont {Paul}}, \bibinfo {author} {\bibfnamefont {A.}~\bibnamefont {Abouelkomsan}}, \bibinfo {author} {\bibfnamefont {A.}~\bibnamefont {Reddy}},\ and\ \bibinfo {author} {\bibfnamefont {L.}~\bibnamefont {Fu}},\ }\bibfield  {title} {\bibinfo {title} {Shining light on collective modes in moir{\'e} fractional {Chern} insulators},\ }\href {https://arxiv.org/abs/2502.17569} {\bibfield  {journal} {\bibinfo  {journal} {arXiv preprint arXiv:2502.17569}\ } (\bibinfo {year} {2025})}\BibitemShut {NoStop}%
\bibitem [{\citenamefont {Mostaan}\ \emph {et~al.}(2025)\citenamefont {Mostaan}, \citenamefont {Goldman}, \citenamefont {{\.I}mamo{\u{g}}lu},\ and\ \citenamefont {Grusdt}}]{mostaan2025anyon}%
  \BibitemOpen
  \bibfield  {author} {\bibinfo {author} {\bibfnamefont {N.}~\bibnamefont {Mostaan}}, \bibinfo {author} {\bibfnamefont {N.}~\bibnamefont {Goldman}}, \bibinfo {author} {\bibfnamefont {A.}~\bibnamefont {{\.I}mamo{\u{g}}lu}},\ and\ \bibinfo {author} {\bibfnamefont {F.}~\bibnamefont {Grusdt}},\ }\bibfield  {title} {\bibinfo {title} {Anyon-trions in atomically thin semiconductor heterostructures},\ }\href {https://arxiv.org/abs/2507.08933} {\bibfield  {journal} {\bibinfo  {journal} {arXiv preprint arXiv:2507.08933}\ } (\bibinfo {year} {2025})}\BibitemShut {NoStop}%
\bibitem [{\citenamefont {He}\ \emph {et~al.}(2015)\citenamefont {He} \emph {et~al.}}]{He:2014bnl}%
  \BibitemOpen
  \bibfield  {author} {\bibinfo {author} {\bibfnamefont {Y.-M.}\ \bibnamefont {He}} \emph {et~al.},\ }\bibfield  {title} {\bibinfo {title} {{Single quantum emitters in monolayer semiconductors}},\ }\href {https://doi.org/10.1038/nnano.2015.75} {\bibfield  {journal} {\bibinfo  {journal} {Nature Nanotech.}\ }\textbf {\bibinfo {volume} {10}},\ \bibinfo {pages} {497} (\bibinfo {year} {2015})},\ \Eprint {https://arxiv.org/abs/1411.2449} {arXiv:1411.2449 [cond-mat.mes-hall]} \BibitemShut {NoStop}%
\bibitem [{\citenamefont {Srivastava}\ \emph {et~al.}(2015)\citenamefont {Srivastava}, \citenamefont {Sidler}, \citenamefont {Allain}, \citenamefont {Lembke}, \citenamefont {Kis},\ and\ \citenamefont {{\.I}mamo{\u{g}}lu}}]{srivastava2015optically}%
  \BibitemOpen
  \bibfield  {author} {\bibinfo {author} {\bibfnamefont {A.}~\bibnamefont {Srivastava}}, \bibinfo {author} {\bibfnamefont {M.}~\bibnamefont {Sidler}}, \bibinfo {author} {\bibfnamefont {A.~V.}\ \bibnamefont {Allain}}, \bibinfo {author} {\bibfnamefont {D.~S.}\ \bibnamefont {Lembke}}, \bibinfo {author} {\bibfnamefont {A.}~\bibnamefont {Kis}},\ and\ \bibinfo {author} {\bibfnamefont {A.}~\bibnamefont {{\.I}mamo{\u{g}}lu}},\ }\bibfield  {title} {\bibinfo {title} {Optically active quantum dots in monolayer {WSe$_2$}},\ }\href {https://doi.org/10.1038/nnano.2015.60} {\bibfield  {journal} {\bibinfo  {journal} {Nature nanotechnology}\ }\textbf {\bibinfo {volume} {10}},\ \bibinfo {pages} {491} (\bibinfo {year} {2015})}\BibitemShut {NoStop}%
\bibitem [{\citenamefont {{Chakraborty}}\ \emph {et~al.}(2015)\citenamefont {{Chakraborty}}, \citenamefont {{Kinnischtzke}}, \citenamefont {{Goodfellow}}, \citenamefont {{Beams}},\ and\ \citenamefont {{Vamivakas}}}]{Chakraborty2015}%
  \BibitemOpen
  \bibfield  {author} {\bibinfo {author} {\bibfnamefont {C.}~\bibnamefont {{Chakraborty}}}, \bibinfo {author} {\bibfnamefont {L.}~\bibnamefont {{Kinnischtzke}}}, \bibinfo {author} {\bibfnamefont {K.~M.}\ \bibnamefont {{Goodfellow}}}, \bibinfo {author} {\bibfnamefont {R.}~\bibnamefont {{Beams}}},\ and\ \bibinfo {author} {\bibfnamefont {A.~N.}\ \bibnamefont {{Vamivakas}}},\ }\bibfield  {title} {\bibinfo {title} {{Voltage-controlled quantum light from an atomically thin semiconductor}},\ }\href {https://doi.org/10.1038/nnano.2015.79} {\bibfield  {journal} {\bibinfo  {journal} {Nature Nanotechnology}\ }\textbf {\bibinfo {volume} {10}},\ \bibinfo {pages} {507} (\bibinfo {year} {2015})}\BibitemShut {NoStop}%
\bibitem [{\citenamefont {Koperski}\ \emph {et~al.}(2015)\citenamefont {Koperski}, \citenamefont {Nogajewski}, \citenamefont {Arora}, \citenamefont {Cherkez}, \citenamefont {Mallet}, \citenamefont {Veuillen}, \citenamefont {Marcus}, \citenamefont {Kossacki},\ and\ \citenamefont {Potemski}}]{koperski2015single}%
  \BibitemOpen
  \bibfield  {author} {\bibinfo {author} {\bibfnamefont {M.}~\bibnamefont {Koperski}}, \bibinfo {author} {\bibfnamefont {K.}~\bibnamefont {Nogajewski}}, \bibinfo {author} {\bibfnamefont {A.}~\bibnamefont {Arora}}, \bibinfo {author} {\bibfnamefont {V.}~\bibnamefont {Cherkez}}, \bibinfo {author} {\bibfnamefont {P.}~\bibnamefont {Mallet}}, \bibinfo {author} {\bibfnamefont {J.-Y.}\ \bibnamefont {Veuillen}}, \bibinfo {author} {\bibfnamefont {J.}~\bibnamefont {Marcus}}, \bibinfo {author} {\bibfnamefont {P.}~\bibnamefont {Kossacki}},\ and\ \bibinfo {author} {\bibfnamefont {M.}~\bibnamefont {Potemski}},\ }\bibfield  {title} {\bibinfo {title} {Single photon emitters in exfoliated {WSe$_2$} structures},\ }\href {https://doi.org/10.1038/nnano.2015.67} {\bibfield  {journal} {\bibinfo  {journal} {Nature nanotechnology}\ }\textbf {\bibinfo {volume} {10}},\ \bibinfo {pages} {503} (\bibinfo {year} {2015})}\BibitemShut {NoStop}%
\bibitem [{\citenamefont {Kumar}\ \emph {et~al.}(2015)\citenamefont {Kumar}, \citenamefont {Kaczmarczyk},\ and\ \citenamefont {Gerardot}}]{Kumar_2015}%
  \BibitemOpen
  \bibfield  {author} {\bibinfo {author} {\bibfnamefont {S.}~\bibnamefont {Kumar}}, \bibinfo {author} {\bibfnamefont {A.}~\bibnamefont {Kaczmarczyk}},\ and\ \bibinfo {author} {\bibfnamefont {B.~D.}\ \bibnamefont {Gerardot}},\ }\bibfield  {title} {\bibinfo {title} {Strain-induced spatial and spectral isolation of quantum emitters in mono- and bilayer {WSe$_2$}},\ }\href {https://doi.org/10.1021/acs.nanolett.5b03312} {\bibfield  {journal} {\bibinfo  {journal} {Nano Letters}\ }\textbf {\bibinfo {volume} {15}},\ \bibinfo {pages} {7567–7573} (\bibinfo {year} {2015})}\BibitemShut {NoStop}%
\bibitem [{\citenamefont {Thureja}(2023)}]{thureja2023electricallytunablequantumconfinement}%
  \BibitemOpen
  \bibfield  {author} {\bibinfo {author} {\bibfnamefont {D.}~\bibnamefont {Thureja}},\ }\href@noop {} {\bibinfo {title} {Electrically tunable quantum confinement of neutral excitons}} (\bibinfo {year} {2023}),\ \Eprint {https://arxiv.org/abs/2308.07879} {arXiv:2308.07879 [cond-mat.mes-hall]} \BibitemShut {NoStop}%
\bibitem [{\citenamefont {Li}\ \emph {et~al.}(2026)\citenamefont {Li}, \citenamefont {Wang~Beach}, \citenamefont {Hu}, \citenamefont {Taniguchi}, \citenamefont {Watanabe}, \citenamefont {Chu}, \citenamefont {Imamo{\u{g}}lu}, \citenamefont {Cao}, \citenamefont {Xiao},\ and\ \citenamefont {Xu}}]{li2026signatures}%
  \BibitemOpen
  \bibfield  {author} {\bibinfo {author} {\bibfnamefont {W.}~\bibnamefont {Li}}, \bibinfo {author} {\bibfnamefont {C.}~\bibnamefont {Wang~Beach}}, \bibinfo {author} {\bibfnamefont {C.}~\bibnamefont {Hu}}, \bibinfo {author} {\bibfnamefont {T.}~\bibnamefont {Taniguchi}}, \bibinfo {author} {\bibfnamefont {K.}~\bibnamefont {Watanabe}}, \bibinfo {author} {\bibfnamefont {J.-H.}\ \bibnamefont {Chu}}, \bibinfo {author} {\bibfnamefont {A.}~\bibnamefont {Imamo{\u{g}}lu}}, \bibinfo {author} {\bibfnamefont {T.}~\bibnamefont {Cao}}, \bibinfo {author} {\bibfnamefont {D.}~\bibnamefont {Xiao}},\ and\ \bibinfo {author} {\bibfnamefont {X.}~\bibnamefont {Xu}},\ }\bibfield  {title} {\bibinfo {title} {Signatures of fractional charges via anyon--trions in twisted {MoTe$_2$}},\ }\href {https://doi.org/10.1038/s41586-026-10101-w} {\bibfield  {journal} {\bibinfo  {journal} {Nature}\ }\textbf {\bibinfo {volume} {651}},\ \bibinfo {pages} {48} (\bibinfo {year} {2026})}\BibitemShut {NoStop}%
\bibitem [{\citenamefont {Rytova}(2018)}]{rytova2018screened}%
  \BibitemOpen
  \bibfield  {author} {\bibinfo {author} {\bibfnamefont {N.~S.}\ \bibnamefont {Rytova}},\ }\bibfield  {title} {\bibinfo {title} {Screened potential of a point charge in a thin film},\ }\href@noop {} {\bibfield  {journal} {\bibinfo  {journal} {arXiv preprint arXiv:1806.00976}\ } (\bibinfo {year} {2018})}\BibitemShut {NoStop}%
\bibitem [{\citenamefont {Keldysh}(1979)}]{Keldysh1979}%
  \BibitemOpen
  \bibfield  {author} {\bibinfo {author} {\bibfnamefont {L.~V.}\ \bibnamefont {Keldysh}},\ }\bibfield  {title} {\bibinfo {title} {Coulomb interaction in thin semiconductor and semimetal films},\ }\href@noop {} {\bibfield  {journal} {\bibinfo  {journal} {JETP Letters}\ }\textbf {\bibinfo {volume} {29}},\ \bibinfo {pages} {658} (\bibinfo {year} {1979})}\BibitemShut {NoStop}%
\bibitem [{\citenamefont {Cudazzo}\ \emph {et~al.}(2011)\citenamefont {Cudazzo}, \citenamefont {Tokatly},\ and\ \citenamefont {Rubio}}]{cudazzo2011dielectric}%
  \BibitemOpen
  \bibfield  {author} {\bibinfo {author} {\bibfnamefont {P.}~\bibnamefont {Cudazzo}}, \bibinfo {author} {\bibfnamefont {I.~V.}\ \bibnamefont {Tokatly}},\ and\ \bibinfo {author} {\bibfnamefont {A.}~\bibnamefont {Rubio}},\ }\bibfield  {title} {\bibinfo {title} {Dielectric screening in two-dimensional insulators: Implications for excitonic and impurity states in graphane},\ }\href {https://doi.org/10.1103/PhysRevB.84.085406} {\bibfield  {journal} {\bibinfo  {journal} {Phys. Rev. B}\ }\textbf {\bibinfo {volume} {84}},\ \bibinfo {pages} {085406} (\bibinfo {year} {2011})}\BibitemShut {NoStop}%
\bibitem [{Note1()}]{Note1}%
  \BibitemOpen
  \bibinfo {note} {Although one anyon is not a strictly speaking gauge invariant object in itself, we will assume that other anyons are present in the sample such that the many-body wave function of the whole system is gauge invariant}\BibitemShut {NoStop}%
\bibitem [{\citenamefont {Mu\~noz de~las Heras}\ \emph {et~al.}(2020)\citenamefont {Mu\~noz de~las Heras}, \citenamefont {Macaluso},\ and\ \citenamefont {Carusotto}}]{Munoz2020}%
  \BibitemOpen
  \bibfield  {author} {\bibinfo {author} {\bibfnamefont {A.}~\bibnamefont {Mu\~noz de~las Heras}}, \bibinfo {author} {\bibfnamefont {E.}~\bibnamefont {Macaluso}},\ and\ \bibinfo {author} {\bibfnamefont {I.}~\bibnamefont {Carusotto}},\ }\bibfield  {title} {\bibinfo {title} {Anyonic molecules in atomic fractional quantum hall liquids: A quantitative probe of fractional charge and anyonic statistics},\ }\href {https://doi.org/10.1103/PhysRevX.10.041058} {\bibfield  {journal} {\bibinfo  {journal} {Phys. Rev. X}\ }\textbf {\bibinfo {volume} {10}},\ \bibinfo {pages} {041058} (\bibinfo {year} {2020})}\BibitemShut {NoStop}%
\bibitem [{\citenamefont {Gattu}\ and\ \citenamefont {Jain}(2025)}]{gattu2025molecular}%
  \BibitemOpen
  \bibfield  {author} {\bibinfo {author} {\bibfnamefont {M.}~\bibnamefont {Gattu}}\ and\ \bibinfo {author} {\bibfnamefont {J.}~\bibnamefont {Jain}},\ }\bibfield  {title} {\bibinfo {title} {Molecular anyons in the fractional quantum hall effect},\ }\href {https://journals.aps.org/prl/abstract/10.1103/PhysRevLett.135.236601} {\bibfield  {journal} {\bibinfo  {journal} {Phys. Rev. Lett.}\ }\textbf {\bibinfo {volume} {135}},\ \bibinfo {pages} {236601} (\bibinfo {year} {2025})}\BibitemShut {NoStop}%
\bibitem [{\citenamefont {Xu}\ \emph {et~al.}(2025{\natexlab{b}})\citenamefont {Xu}, \citenamefont {Ji}, \citenamefont {Wang}, \citenamefont {Trung},\ and\ \citenamefont {Yang}}]{Yang2025}%
  \BibitemOpen
  \bibfield  {author} {\bibinfo {author} {\bibfnamefont {Q.}~\bibnamefont {Xu}}, \bibinfo {author} {\bibfnamefont {G.}~\bibnamefont {Ji}}, \bibinfo {author} {\bibfnamefont {Y.}~\bibnamefont {Wang}}, \bibinfo {author} {\bibfnamefont {H.~Q.}\ \bibnamefont {Trung}},\ and\ \bibinfo {author} {\bibfnamefont {B.}~\bibnamefont {Yang}},\ }\bibfield  {title} {\bibinfo {title} {Dynamics of anyon clusters in fractional quantum hall fluids},\ }\href {https://doi.org/10.1103/vgz6-z98r} {\bibfield  {journal} {\bibinfo  {journal} {Phys. Rev. B}\ }\textbf {\bibinfo {volume} {112}},\ \bibinfo {pages} {235112} (\bibinfo {year} {2025}{\natexlab{b}})}\BibitemShut {NoStop}%
\bibitem [{\citenamefont {Laughlin}(1984)}]{LAUGHLIN1984}%
  \BibitemOpen
  \bibfield  {author} {\bibinfo {author} {\bibfnamefont {R.}~\bibnamefont {Laughlin}},\ }\bibfield  {title} {\bibinfo {title} {Excitons in the fractional quantum hall effect},\ }\href {https://doi.org/https://doi.org/10.1016/0378-4363(84)90172-4} {\bibfield  {journal} {\bibinfo  {journal} {Physica B+C}\ }\textbf {\bibinfo {volume} {126}},\ \bibinfo {pages} {254} (\bibinfo {year} {1984})}\BibitemShut {NoStop}%
\bibitem [{\citenamefont {Haldane}\ and\ \citenamefont {Rezayi}(1985)}]{Haldane1985}%
  \BibitemOpen
  \bibfield  {author} {\bibinfo {author} {\bibfnamefont {F.~D.~M.}\ \bibnamefont {Haldane}}\ and\ \bibinfo {author} {\bibfnamefont {E.~H.}\ \bibnamefont {Rezayi}},\ }\bibfield  {title} {\bibinfo {title} {Finite-size studies of the incompressible state of the fractionally quantized hall effect and its excitations},\ }\href {https://doi.org/10.1103/PhysRevLett.54.237} {\bibfield  {journal} {\bibinfo  {journal} {Phys. Rev. Lett.}\ }\textbf {\bibinfo {volume} {54}},\ \bibinfo {pages} {237} (\bibinfo {year} {1985})}\BibitemShut {NoStop}%
\bibitem [{\citenamefont {Kamilla}\ \emph {et~al.}(1996)\citenamefont {Kamilla}, \citenamefont {Wu},\ and\ \citenamefont {Jain}}]{Kamilla1996}%
  \BibitemOpen
  \bibfield  {author} {\bibinfo {author} {\bibfnamefont {R.~K.}\ \bibnamefont {Kamilla}}, \bibinfo {author} {\bibfnamefont {X.~G.}\ \bibnamefont {Wu}},\ and\ \bibinfo {author} {\bibfnamefont {J.~K.}\ \bibnamefont {Jain}},\ }\bibfield  {title} {\bibinfo {title} {Excitons of composite fermions},\ }\href {https://doi.org/10.1103/PhysRevB.54.4873} {\bibfield  {journal} {\bibinfo  {journal} {Phys. Rev. B}\ }\textbf {\bibinfo {volume} {54}},\ \bibinfo {pages} {4873} (\bibinfo {year} {1996})}\BibitemShut {NoStop}%
\bibitem [{\citenamefont {Yang}\ \emph {et~al.}(2012)\citenamefont {Yang}, \citenamefont {Hu}, \citenamefont {Papi\ifmmode~\acute{c}\else \'{c}\fi{}},\ and\ \citenamefont {Haldane}}]{Yang2012}%
  \BibitemOpen
  \bibfield  {author} {\bibinfo {author} {\bibfnamefont {B.}~\bibnamefont {Yang}}, \bibinfo {author} {\bibfnamefont {Z.-X.}\ \bibnamefont {Hu}}, \bibinfo {author} {\bibfnamefont {Z.}~\bibnamefont {Papi\ifmmode~\acute{c}\else \'{c}\fi{}}},\ and\ \bibinfo {author} {\bibfnamefont {F.~D.~M.}\ \bibnamefont {Haldane}},\ }\bibfield  {title} {\bibinfo {title} {Model wave functions for the collective modes and the magnetoroton theory of the fractional quantum hall effect},\ }\href {https://doi.org/10.1103/PhysRevLett.108.256807} {\bibfield  {journal} {\bibinfo  {journal} {Phys. Rev. Lett.}\ }\textbf {\bibinfo {volume} {108}},\ \bibinfo {pages} {256807} (\bibinfo {year} {2012})}\BibitemShut {NoStop}%
\bibitem [{\citenamefont {Jolicoeur}(2017)}]{PhysRevB.95.075201}%
  \BibitemOpen
  \bibfield  {author} {\bibinfo {author} {\bibfnamefont {T.}~\bibnamefont {Jolicoeur}},\ }\bibfield  {title} {\bibinfo {title} {Shape of the magnetoroton at $\ensuremath{\nu}=1/3$ and $\ensuremath{\nu}=7/3$ in real samples},\ }\href {https://doi.org/10.1103/PhysRevB.95.075201} {\bibfield  {journal} {\bibinfo  {journal} {Phys. Rev. B}\ }\textbf {\bibinfo {volume} {95}},\ \bibinfo {pages} {075201} (\bibinfo {year} {2017})}\BibitemShut {NoStop}%
\bibitem [{\citenamefont {Suzuki}\ and\ \citenamefont {Varga}(2002)}]{suzuki2002stochastic}%
  \BibitemOpen
  \bibfield  {author} {\bibinfo {author} {\bibfnamefont {Y.}~\bibnamefont {Suzuki}}\ and\ \bibinfo {author} {\bibfnamefont {K.}~\bibnamefont {Varga}},\ }\href {https://link.springer.com/book/9783540651529} {\emph {\bibinfo {title} {Stochastic variational approach to quantum-mechanical few-body problems}}}\ (\bibinfo  {publisher} {Springer},\ \bibinfo {year} {2002})\BibitemShut {NoStop}%
\bibitem [{\citenamefont {Mitroy}\ \emph {et~al.}(2013)\citenamefont {Mitroy}, \citenamefont {Bubin}, \citenamefont {Horiuchi}, \citenamefont {Suzuki}, \citenamefont {Adamowicz}, \citenamefont {Cencek}, \citenamefont {Szalewicz}, \citenamefont {Komasa}, \citenamefont {Blume},\ and\ \citenamefont {Varga}}]{mitroy2013theory}%
  \BibitemOpen
  \bibfield  {author} {\bibinfo {author} {\bibfnamefont {J.}~\bibnamefont {Mitroy}}, \bibinfo {author} {\bibfnamefont {S.}~\bibnamefont {Bubin}}, \bibinfo {author} {\bibfnamefont {W.}~\bibnamefont {Horiuchi}}, \bibinfo {author} {\bibfnamefont {Y.}~\bibnamefont {Suzuki}}, \bibinfo {author} {\bibfnamefont {L.}~\bibnamefont {Adamowicz}}, \bibinfo {author} {\bibfnamefont {W.}~\bibnamefont {Cencek}}, \bibinfo {author} {\bibfnamefont {K.}~\bibnamefont {Szalewicz}}, \bibinfo {author} {\bibfnamefont {J.}~\bibnamefont {Komasa}}, \bibinfo {author} {\bibfnamefont {D.}~\bibnamefont {Blume}},\ and\ \bibinfo {author} {\bibfnamefont {K.}~\bibnamefont {Varga}},\ }\bibfield  {title} {\bibinfo {title} {Theory and application of explicitly correlated gaussians},\ }\href {https://journals.aps.org/rmp/abstract/10.1103/RevModPhys.85.693} {\bibfield  {journal} {\bibinfo  {journal} {Reviews of Modern Physics}\ }\textbf {\bibinfo {volume} {85}},\ \bibinfo {pages} {693} (\bibinfo {year} {2013})}\BibitemShut {NoStop}%
\bibitem [{\citenamefont {Katow}\ \emph {et~al.}(2017)\citenamefont {Katow}, \citenamefont {Usukura}, \citenamefont {Akashi}, \citenamefont {Varga},\ and\ \citenamefont {Tsuneyuki}}]{katow2017numerical}%
  \BibitemOpen
  \bibfield  {author} {\bibinfo {author} {\bibfnamefont {H.}~\bibnamefont {Katow}}, \bibinfo {author} {\bibfnamefont {J.}~\bibnamefont {Usukura}}, \bibinfo {author} {\bibfnamefont {R.}~\bibnamefont {Akashi}}, \bibinfo {author} {\bibfnamefont {K.}~\bibnamefont {Varga}},\ and\ \bibinfo {author} {\bibfnamefont {S.}~\bibnamefont {Tsuneyuki}},\ }\bibfield  {title} {\bibinfo {title} {Numerical investigation of triexciton stabilization in diamond with multiple valleys and bands},\ }\href {https://journals.aps.org/prb/abstract/10.1103/PhysRevB.95.125205} {\bibfield  {journal} {\bibinfo  {journal} {Phys. Rev. B}\ }\textbf {\bibinfo {volume} {95}},\ \bibinfo {pages} {125205} (\bibinfo {year} {2017})}\BibitemShut {NoStop}%
\bibitem [{\citenamefont {Van~der Donck}\ \emph {et~al.}(2017)\citenamefont {Van~der Donck}, \citenamefont {Zarenia},\ and\ \citenamefont {Peeters}}]{van2017excitons}%
  \BibitemOpen
  \bibfield  {author} {\bibinfo {author} {\bibfnamefont {M.}~\bibnamefont {Van~der Donck}}, \bibinfo {author} {\bibfnamefont {M.}~\bibnamefont {Zarenia}},\ and\ \bibinfo {author} {\bibfnamefont {F.}~\bibnamefont {Peeters}},\ }\bibfield  {title} {\bibinfo {title} {Excitons and trions in monolayer transition metal dichalcogenides: A comparative study between the multiband model and the quadratic single-band model},\ }\href {https://journals.aps.org/prb/abstract/10.1103/PhysRevB.96.035131} {\bibfield  {journal} {\bibinfo  {journal} {Phys. Rev. B}\ }\textbf {\bibinfo {volume} {96}},\ \bibinfo {pages} {035131} (\bibinfo {year} {2017})}\BibitemShut {NoStop}%
\bibitem [{\citenamefont {Van~der Donck}\ \emph {et~al.}(2018)\citenamefont {Van~der Donck}, \citenamefont {Zarenia},\ and\ \citenamefont {Peeters}}]{van2018excitons}%
  \BibitemOpen
  \bibfield  {author} {\bibinfo {author} {\bibfnamefont {M.}~\bibnamefont {Van~der Donck}}, \bibinfo {author} {\bibfnamefont {M.}~\bibnamefont {Zarenia}},\ and\ \bibinfo {author} {\bibfnamefont {F.}~\bibnamefont {Peeters}},\ }\bibfield  {title} {\bibinfo {title} {Excitons, trions, and biexcitons in transition-metal dichalcogenides: magnetic-field dependence},\ }\href {https://journals.aps.org/prb/abstract/10.1103/PhysRevB.97.195408} {\bibfield  {journal} {\bibinfo  {journal} {Phys. Rev. B}\ }\textbf {\bibinfo {volume} {97}},\ \bibinfo {pages} {195408} (\bibinfo {year} {2018})}\BibitemShut {NoStop}%
\bibitem [{\citenamefont {Cho}\ \emph {et~al.}(2021)\citenamefont {Cho}, \citenamefont {Greene},\ and\ \citenamefont {Berkelbach}}]{cho2021simulations}%
  \BibitemOpen
  \bibfield  {author} {\bibinfo {author} {\bibfnamefont {Y.}~\bibnamefont {Cho}}, \bibinfo {author} {\bibfnamefont {S.~M.}\ \bibnamefont {Greene}},\ and\ \bibinfo {author} {\bibfnamefont {T.~C.}\ \bibnamefont {Berkelbach}},\ }\bibfield  {title} {\bibinfo {title} {Simulations of trions and biexcitons in layered hybrid organic-inorganic lead halide perovskites},\ }\href {https://journals.aps.org/prl/abstract/10.1103/PhysRevLett.126.216402} {\bibfield  {journal} {\bibinfo  {journal} {Phys. Rev. Lett.}\ }\textbf {\bibinfo {volume} {126}},\ \bibinfo {pages} {216402} (\bibinfo {year} {2021})}\BibitemShut {NoStop}%
\bibitem [{\citenamefont {Ten{\'o}rio}\ \emph {et~al.}(2026)\citenamefont {Ten{\'o}rio}, \citenamefont {Chaves}, \citenamefont {Hiyama},\ and\ \citenamefont {Frederico}}]{tenorio2026gaussian}%
  \BibitemOpen
  \bibfield  {author} {\bibinfo {author} {\bibfnamefont {L.~G.}\ \bibnamefont {Ten{\'o}rio}}, \bibinfo {author} {\bibfnamefont {A.~J.}\ \bibnamefont {Chaves}}, \bibinfo {author} {\bibfnamefont {E.}~\bibnamefont {Hiyama}},\ and\ \bibinfo {author} {\bibfnamefont {T.}~\bibnamefont {Frederico}},\ }\bibfield  {title} {\bibinfo {title} {Gaussian expansion method for few-body states in two-dimensional materials},\ }\href {https://arxiv.org/abs/2602.11386} {\bibfield  {journal} {\bibinfo  {journal} {arXiv preprint arXiv:2602.11386}\ } (\bibinfo {year} {2026})}\BibitemShut {NoStop}%
\bibitem [{Note2()}]{Note2}%
  \BibitemOpen
  \bibinfo {note} {We used the definition $r_T^2=\protect \frac {1}{3}\expval {(r_{h_1h_2}^2+r_{eh_1}^2+r_{eh_2}^2)}$.}\BibitemShut {Stop}%
\bibitem [{\citenamefont {{Reddy}}\ \emph {et~al.}(2023)\citenamefont {{Reddy}}, \citenamefont {{Alsallom}}, \citenamefont {{Zhang}}, \citenamefont {{Devakul}},\ and\ \citenamefont {{Fu}}}]{Reddy2023}%
  \BibitemOpen
  \bibfield  {author} {\bibinfo {author} {\bibfnamefont {A.~P.}\ \bibnamefont {{Reddy}}}, \bibinfo {author} {\bibfnamefont {F.}~\bibnamefont {{Alsallom}}}, \bibinfo {author} {\bibfnamefont {Y.}~\bibnamefont {{Zhang}}}, \bibinfo {author} {\bibfnamefont {T.}~\bibnamefont {{Devakul}}},\ and\ \bibinfo {author} {\bibfnamefont {L.}~\bibnamefont {{Fu}}},\ }\bibfield  {title} {\bibinfo {title} {{Fractional quantum anomalous Hall states in twisted bilayer MoTe$_{2}$ and WSe$_{2}$}},\ }\href {https://doi.org/10.1103/PhysRevB.108.085117} {\bibfield  {journal} {\bibinfo  {journal} {\prb}\ }\textbf {\bibinfo {volume} {108}},\ \bibinfo {eid} {085117} (\bibinfo {year} {2023})},\ \Eprint {https://arxiv.org/abs/2304.12261} {arXiv:2304.12261 [cond-mat.mes-hall]} \BibitemShut {NoStop}%
\bibitem [{\citenamefont {{Reddy}}\ and\ \citenamefont {{Fu}}(2023)}]{Reddy2023a}%
  \BibitemOpen
  \bibfield  {author} {\bibinfo {author} {\bibfnamefont {A.~P.}\ \bibnamefont {{Reddy}}}\ and\ \bibinfo {author} {\bibfnamefont {L.}~\bibnamefont {{Fu}}},\ }\bibfield  {title} {\bibinfo {title} {{Toward a global phase diagram of the fractional quantum anomalous Hall effect}},\ }\href {https://doi.org/10.1103/PhysRevB.108.245159} {\bibfield  {journal} {\bibinfo  {journal} {\prb}\ }\textbf {\bibinfo {volume} {108}},\ \bibinfo {eid} {245159} (\bibinfo {year} {2023})},\ \Eprint {https://arxiv.org/abs/2308.10406} {arXiv:2308.10406 [cond-mat.mes-hall]} \BibitemShut {NoStop}%
\bibitem [{\citenamefont {Wagner}\ \emph {et~al.}(2026)\citenamefont {Wagner}, \citenamefont {Adlong}, \citenamefont {Christianen},\ and\ \citenamefont {Imamoglu}}]{Wagner2026}%
  \BibitemOpen
  \bibfield  {author} {\bibinfo {author} {\bibfnamefont {G.}~\bibnamefont {Wagner}}, \bibinfo {author} {\bibfnamefont {H.}~\bibnamefont {Adlong}}, \bibinfo {author} {\bibfnamefont {A.}~\bibnamefont {Christianen}},\ and\ \bibinfo {author} {\bibfnamefont {A.}~\bibnamefont {Imamoglu}},\ }\href@noop {} {\bibinfo {title} {in preparation}} (\bibinfo {year} {2026})\BibitemShut {NoStop}%
\bibitem [{\citenamefont {Arovas}\ \emph {et~al.}(1984)\citenamefont {Arovas}, \citenamefont {Schrieffer},\ and\ \citenamefont {Wilczek}}]{Arovas1984FQHStatistics}%
  \BibitemOpen
  \bibfield  {author} {\bibinfo {author} {\bibfnamefont {D.}~\bibnamefont {Arovas}}, \bibinfo {author} {\bibfnamefont {J.~R.}\ \bibnamefont {Schrieffer}},\ and\ \bibinfo {author} {\bibfnamefont {F.}~\bibnamefont {Wilczek}},\ }\bibfield  {title} {\bibinfo {title} {Fractional statistics and the quantum hall effect},\ }\href {https://doi.org/10.1103/PhysRevLett.53.722} {\bibfield  {journal} {\bibinfo  {journal} {Phys. Rev. Lett.}\ }\textbf {\bibinfo {volume} {53}},\ \bibinfo {pages} {722} (\bibinfo {year} {1984})}\BibitemShut {NoStop}%
\bibitem [{\citenamefont {Redekop}\ \emph {et~al.}(2024)\citenamefont {Redekop}, \citenamefont {Zhang}, \citenamefont {Park}, \citenamefont {Cai}, \citenamefont {Anderson}, \citenamefont {Sheekey}, \citenamefont {Arp}, \citenamefont {Babikyan}, \citenamefont {Salters}, \citenamefont {Watanabe}, \citenamefont {Taniguchi}, \citenamefont {Crommie}, \citenamefont {Zaletel},\ and\ \citenamefont {Wang}}]{redekop2024direct}%
  \BibitemOpen
  \bibfield  {author} {\bibinfo {author} {\bibfnamefont {E.}~\bibnamefont {Redekop}}, \bibinfo {author} {\bibfnamefont {C.}~\bibnamefont {Zhang}}, \bibinfo {author} {\bibfnamefont {H.}~\bibnamefont {Park}}, \bibinfo {author} {\bibfnamefont {J.}~\bibnamefont {Cai}}, \bibinfo {author} {\bibfnamefont {E.}~\bibnamefont {Anderson}}, \bibinfo {author} {\bibfnamefont {O.}~\bibnamefont {Sheekey}}, \bibinfo {author} {\bibfnamefont {T.}~\bibnamefont {Arp}}, \bibinfo {author} {\bibfnamefont {G.}~\bibnamefont {Babikyan}}, \bibinfo {author} {\bibfnamefont {S.}~\bibnamefont {Salters}}, \bibinfo {author} {\bibfnamefont {K.}~\bibnamefont {Watanabe}}, \bibinfo {author} {\bibfnamefont {T.}~\bibnamefont {Taniguchi}}, \bibinfo {author} {\bibfnamefont {M.~F.}\ \bibnamefont {Crommie}}, \bibinfo {author} {\bibfnamefont {M.~P.}\ \bibnamefont {Zaletel}},\ and\ \bibinfo {author} {\bibfnamefont {F.}~\bibnamefont {Wang}},\ }\bibfield  {title} {\bibinfo {title} {Direct magnetic imaging of fractional chern insulators in twisted {MoTe$_2$}
  with a superconducting sensor},\ }\href {https://doi.org/10.1038/s41586-024-08153-x} {\bibfield  {journal} {\bibinfo  {journal} {Nature}\ }\textbf {\bibinfo {volume} {635}},\ \bibinfo {pages} {584} (\bibinfo {year} {2024})}\BibitemShut {NoStop}%
\end{thebibliography}%

\clearpage 
\onecolumngrid 
\appendix

\section{Numerical methods}\label{app:Numerical Methods}

Our model for the anyon-trion system is a few-body quantum mechanical problem described by the Hamiltonian
\begin{equation}
    H = \sum_{j} \left(\frac{\boldsymbol{p}_j^2}{2m_j} + q_j U(\boldsymbol{r}_j) \right)+ \sum_{i<j} q_i q_j V^{\text{KR}}(\abs{\boldsymbol{r}_i-\boldsymbol{r}_j})\,,
\end{equation}
with indices $i,j$ labeling the various particles. For the anyon-trion these would be the two holes, $h_1$ and $h_2$, the electron $e$ and the anyon $a$. $V^{\text{KR}}$ is the Keldysh-Rytova potential, given by,
\begin{equation}
    V^\text{KR}(r) =  \frac{\pi e^2}{2\kappa\, r_0}\frac{1}{4\pi\varepsilon_0}\Big[\StruveH_0(r/r_0)- \BesselY_0\left(r/r_0\right)\Big]\,,
\end{equation}
where $\StruveH_0$ and $\BesselY_0$ are the zeroth Struve H function and the Bessel function of second kind, respectively. For the numerics, we set the electron and hole masses to their monolayer MoTe$_2$ values, $m_e = m_h = 0.65 m_0$, where $m_0$ is the bare electron mass. We choose $\kappa=4$ for the dielectric constant and $r_0 = 3.17$nm for the screening length in the Keldysh-Rytova interaction.

Since the anyon braids trivially with both holes as well as the electron, no additional statistical gauge field is required to implement fractional braiding statistics, and the Keldysh-Rytova potential characterizes all interactions between the various particles.

\begin{figure}[t]
    \centering
    \includegraphics[width=0.5\columnwidth]{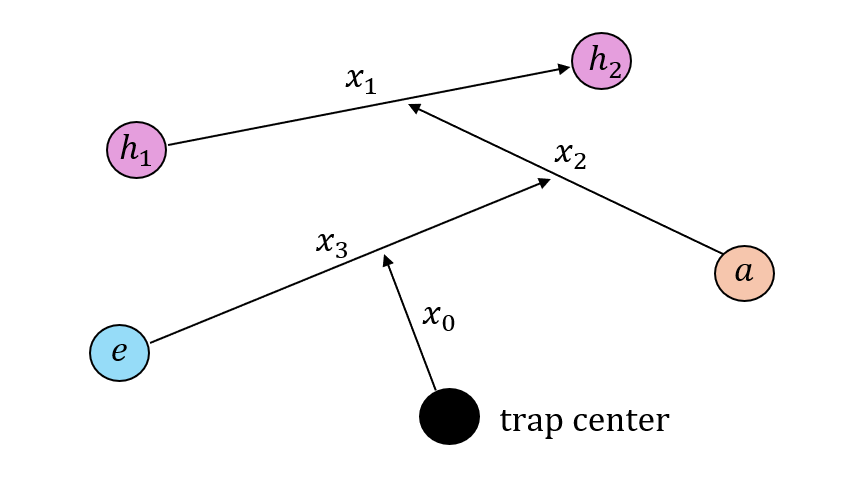}
    \caption{Schematic illustration of the center of mass coordinate $\vec{x}_0$ and three Jacobi coordinates $\vec{x}_1, \vec{x}_2, \vec{x}_3$ for the anyon-trion system.}
    \label{fig:Jacobi_coords}
\end{figure}

We seek the ground state energy $E_\mathrm{min}$ of this Hamiltonian using a variational algorithm. While the presence of the trapping potential $U(\boldsymbol{r})$ prevents us from reducing the dimension of the problem by decoupling the center of mass (COM) coordinate, we nevertheless separate the COM coordinate in order to study various bound states in the absence of the trap. We use Jacobi coordinates for the remaining degrees of freedom (see Fig.~\ref{fig:Jacobi_coords}),

\begin{align}
    \vec{x}_0&=\frac{1}{2m_h+m_e+m_a}(m_h\boldsymbol{r}_{h_1}+m_h\boldsymbol{r}_{h_2}+m_e\boldsymbol{r}_e+m_a\boldsymbol{r}_a)\,,\nonumber \\
    \vec{x}_1&=\boldsymbol{r}_{h_1}-\boldsymbol{r}_{h_2}\,, \qquad
    \vec{x}_2= \boldsymbol{r}_a-\frac{\boldsymbol{r}_{h_1}+\boldsymbol{r}_{h_2}}{2}\,, \qquad
    \vec{x}_3=\boldsymbol{r}_e-\frac{m_h\boldsymbol{r}_{h_1}+m_h\boldsymbol{r}_{h_2}+m_a\boldsymbol{r}_a}{2m_h+m_a}\,.
\end{align}

The Hamiltonian takes the following form in these coordinates:
\begin{align}
    H=\sum_{I=0}^3\frac{\boldsymbol{p}_I^2}{2\mu_I}+\sum_i q_i\, U(\boldsymbol{r}_i)+\sum_{i<j}q_i\,q_j\,V^{\text{KR}}(|\boldsymbol{r}_i-\boldsymbol{r}_j|)\,,
\end{align}
where $\boldsymbol{p}_I=\mu_{I}\dot{\vec{x}}_I$, and the reduced masses are given by
\begin{align}
    \mu_0=2m_h+m_e+m_a\,, \quad
    \mu_1=\frac{m_h}{2}\,, \quad \nonumber\\
    \mu_2=\frac{2m_h\,m_a}{2m_h+m_a}\,, \quad
    \mu_3=\frac{(2m_h+m_a)m_e}{2m_h+m_a+m_e}\,.
\end{align}

We use an explicitly correlated Gaussian (ECG) basis $\phi_\alpha$ for the variational algorithm. Since we expect the ground state to not possess any angular momentum, and every particle to also live in rotationally invariant partial wavefunction, we will use the minimal ECG basis consisting only of exponential, rotationally invariant functions. Additionally, the two holes in the anyon-trion are drawn from different valleys, and therefore carry opposite spins due to spin-valley locking in $t$MoTe$_2$. They form a spin singlet, so that the real space wavefunction should then be symmetrized in $h_1$ and $h_2$. The following basis is best suited for the calculation:
\begin{align}
    \phi_{\alpha}=&\exp \Big(-\sum_{I=0}^3\, \vec{x}_I\, M^\alpha_{IJ}\, \vec{x}_J\Big)+\exp \Big(-\sum_{I=0}^3\, P_{IL}\vec{x}_L\, M^\alpha_{IJ}\, P_{JK}\vec{x}_K\Big)\,,
    \label{appendix:phi symmetrized}
\end{align}
where $P_{JK} = \mathrm{diag}(1,-1,1,1)$. When acting upon $\vec{x}_J$, $P_{IJ}$ flips the sign of $\vec{x}_1$, thereby exchanging the two holes. The numbers $M^\alpha_{IJ}$ are $4\times 4$ matrices indexed by a label $\alpha = 1,\ldots,\mathcal{N}$, where $\mathcal{N}$ is the basis size. These can be assembled into a large $4\mathcal{N}\times 4\mathcal{N}$ block-diagonal matrix with $4\times 4$ blocks.

Let $\psi$ be the variational ground state of the Hamiltonian $H$. We can expand this in the ECG basis by writing $\psi = \sum_{\alpha=1}^\mathcal{N} C_\alpha \phi_\alpha$. The eigenvalue equation for the ground state with energy $E_\mathrm{min}$ then becomes
\begin{align}
   \sum_{\beta=1}^{\mathcal{N}} \langle \phi_{\alpha}|H|\phi_{\beta}\rangle C_{\beta}=E_{\text{min}}\sum_{\beta=1}^{\mathcal{N}}\langle \phi_{\alpha}|\phi_{\beta}\rangle C_{\beta}\,.
   \label{appendix: generalized eigenvalue equation}
\end{align}

The energy $E_\mathrm{min}$ is computed via the stochastic variational method~\cite{suzuki2002stochastic}, which consists of a sequence of runs with increasing basis size $\mathcal{N}$. In the first run, we set $\mathcal{N}=1$ and randomly generate a large enough number (in our case, 2000) of candidate $M^\alpha$ matrices. We then pick the matrix $M^1$ that supplies the lowest eigenvalue $E_\mathrm{min}$ and add the corresponding $\phi_\alpha$ to the basis set. Denoting this basis function as $\phi_1$, the basis set is now $\{ \phi_1 \}$.

The next run sets $\mathcal{N}=2$, and we randomly generate enough new $M^\alpha$'s, each of which is combined with $\phi_1$ to form a two-dimensional basis to solve Eq.~\eqref{appendix: generalized eigenvalue equation}. The matrix $M^2$ supplying the lowest eigenvalue will be kept, and the others discarded, to expand the basis set to $\{ \phi_1, \phi_2 \}$.

This process repeats until $E_\mathrm{min}$ converges to required precision. We do this $150$ times, namely $\mathcal{N}_\mathrm{final} = 150$, at which point fluctuations in $E_\mathrm{min}$ are $\mathcal{O}(0.01)$ meV.

The various matrix elements on the right-hand side of Eq.~\eqref{appendix: generalized eigenvalue equation} can be analytically computed, and we direct the interested reader to Ref.~\cite{mitroy2013theory} and the references therein for more general cases. For the overlap matrix, we have
\begin{align}
    \langle \phi_\alpha|\phi_\beta\rangle
    &=\int \phi_{\alpha}^{\dagger}\phi_\beta\,\prod_{I=0}^3d^2x_I \nonumber\\
    &=\frac{\pi^4}{\det(M^\alpha+M^\beta)}
    +\frac{\pi^4}{\det (P M^\alpha P+M^\beta)}
    +\frac{\pi^4}{\det(M^\alpha+P M^\beta P)}
    +\frac{\pi^4}{\det (P M^\alpha P+P M^\beta P)}\,.
\end{align}
where $P$ is the symmetrization matrix. We will omit the symmetrization for the remaining expressions for the sake of simplicity. The results for the symmetrized basis elements can be recovered by adding three additional terms obtain by replacing $M^\alpha \rightarrow P M^\alpha P$, $M^\beta\rightarrow P M^\beta P$, and both. The kinetic energy takes the form
\begin{align}
    &-\frac{1}{2\mu_I} \langle \phi_{\alpha}|\vec{\nabla}_I^2|\phi_{\beta}\rangle\nonumber\\
    =&\frac{2M^\alpha_{IJ}M^\beta_{IK}}{\mu_I}\int_x\, \vec{x}_J^T\cdot  \vec{x}_K\, \exp\Big(- \sum_{L,M}\vec{x}_L\, (M^\alpha+M^\beta)_{LM}\,\vec{x}_M\Big) \nonumber\\
    =&\frac{2}{\mu_I}\frac{\pi^4}{\det(M^\alpha+M^\beta)}\sum_{J,K}\Big(M^\alpha_{IJ}\, (M^\alpha+M^\beta)^{-1}_{JK}\,M^\beta_{IK}\Big)\,.
\end{align}
Here we abbreviated $\int\prod_{I}d^2x_I$ to $\int_x$.

The Keldysh-Rytova interaction looks much simpler in momentum space,
\begin{align}
    V^{\text{KR}}(p)=\int d^2x\, V^{\text{KR}}(x)\, e^{i\vec{p}\cdot \vec{x}}=\frac{q_iq_je^2}{4\pi \epsilon_0 \kappa}\frac{2\pi}{ p(1+ p\,r_0)}\,.
\end{align}
where $p=|\boldsymbol{p}|$. We compute the matrix elements of the anyon-electron ($ae$) interaction as an example. In the Jacobi basis, we have,
\begin{align}
    &\vec r_a-\vec r_e = -\frac{2m_h}{2m_h+m_a}\,\vec x_2 + \vec x_3 = \sum_{I=0}^3 w^{ae}_I\, \vec{x}_I\,, \qquad w_I^{ae}=(0,0,-\frac{2m_h}{2m_h+m_a},1)\,,
\end{align}
The ECG basis functions Fourier transform to
\begin{align}
    \int_x \exp\!\Big(-\sum_{L,M}&\vec x_L M^\alpha_{LM}\vec x_M\Big)\,
    e^{i\vec p\cdot \sum_I w_I \vec x_I} = \frac{\pi^4}{\det M^\alpha}e^{-\frac12 s p^2}\,, \qquad s=\frac{1}{2}\, \sum_{I,J}w_I[(M^\alpha)^{-1}]_{IJ}w_J\,.
\end{align}

Therefore,
\begin{align}
    &\langle \phi_\alpha|V^{\text{KR}}(|\boldsymbol{r}_a-\boldsymbol{r}_e|)|\phi_\beta\rangle\nonumber\\
    =&\frac{\pi^4}{\det(M^\alpha+M^\beta)}
    \int\frac{d^2p}{(2\pi)^2}\,V^{\text{KR}}(p)\,
    \exp\!\Big(-\frac12 s^{ae} p^2\Big) \nonumber\\
    =&\frac{\pi^4}{\det(M^\alpha+M^\beta)}\,
    \frac{q_aq_ee^2}{4\pi \epsilon_0 \kappa}\, I^{\mathrm{KR}}(s^{ae},r_0)\,.
\end{align}
When $r_0>0$, one has
\begin{equation}
    I^{\mathrm{KR}}(u,r_0)=
    \frac{2\sqrt{\pi}\,D(u)-e^{-u^2}\,\mathrm{Ei}(u^2)}{2r_0}\,,
\end{equation}
where $D(u)$ is Dawson function and $\mathrm{Ei}(u^2)$ is exponential integral, which can both be evaluated efficiently by the standard special function libraries of, for example, \texttt{Julia}. In the limit $r_0\to 0$ where the Keldysh-Rytova potential reduces to the Coulomb potential, one has 
\begin{equation}
    \lim_{r_0\to 0}I^{\mathrm{KR}}(s^{ae},r_0)=\int_0^\infty e^{-\frac12 s^{ae} p^2}\,dp
    =\sqrt{\frac{\pi}{2s^{ae}}}\,.
\end{equation}
The Keldysh-Rytova interaction between other pairs can be easily included by replacing $w^{ae}$ (or $s^{ae}$) with the corresponding ones.

Lastly, the matrix elements of the Gaussian trap potential can also be similarly evaluated. For the anyon, we have
\begin{align}
    &\boldsymbol{r}_a=\sum_I w^a_I\,\vec{x}_I\, , \qquad w^a_I=\left( 1,0,\frac{2m_h}{2m_h+m_a},-\frac{m_e}{2m_h+m_e+m_a} \right)\,.
\end{align}
In momentum space, the matrix elements take the form 
\begin{align}
    &q_a\langle \phi_\alpha |U(\boldsymbol{p}_a)|\phi_{\beta}\rangle\nonumber\\
    =&\frac{q_a2\pi^5\sigma^2 U_0}{\det(M^\alpha+M^\beta)}
    \int \frac{d^2p}{(2\pi)^2}\,
    \exp\!\Big[-\frac12 (s^{a}+\sigma^2) p^2\Big] \nonumber\\
    =&q_aU_0\frac{\pi^4}{\det(M^\alpha+M^\beta)}\frac{\sigma^2}{\sigma^2+s^a}\,.
\end{align}
where $s^a=w^a_I[(M^\alpha+M^\beta)^{-1}]_{IJ}w_J^a/2$.

\onecolumngrid 
\section{\label{app: fractional exciton energies}Fractional exciton energies}

Our model can be generalized to include the energetic effects of braiding between multiple anyons as well, following Ref.~\cite{Arovas1984FQHStatistics}.
This is done by introducing a statistical gauge field $\vec{a}$ with no dynamics that couples only to fractional excitations. The Hamiltonian takes the form
\begin{align}
    H= -\sum_{i}\frac{\vec{D}_i^2}{2m_a} + \sum_{i<j} V^{\text{KR}}(|\boldsymbol{r}_i-\boldsymbol{r}_j|)\,,
\end{align}
where $\vec{D}_i=\vec{\nabla}_i+i\theta q_i \vec{a}(\boldsymbol{r}_i)$ and $\theta$ is the statistical angle of the anyon and $q_i=+1$ for anyon (same sign with hole) and $-1$ for anti-anyon. The statistical gauge field is set to have the following configuration:
\begin{align}
    \vec{a}(\boldsymbol{r}_i)=\sum_{j\neq i}q_j\frac{\hat{z}\times (\boldsymbol{r}_{i}-\boldsymbol{r}_j)}{|\boldsymbol{r}_{i}-\boldsymbol{r}_j|^2}\,.
\end{align}
In other words, the statistical gauge field is a means to attach flux to each anyon, whose Aharanov-Bohm phase implements fractional braiding statistics. The kinetic energy term expands to the following:
\begin{align}\label{eq:H:fewbdy:Statistical}
    -\sum_{i=1}^N\frac{\vec{D}_i^2}{2m_a}
    =\sum_{i=1}^N\frac{-\nabla_i^2}{2m_a}
    -i\frac{\theta}{m_a}\sum_{j\neq i}q_jq_j\frac{(\boldsymbol{r}_{ij}\times \nabla_i)_z}{|\boldsymbol{r}_{ij}|^2}
    +\frac{\theta^2}{2m_a}\sum_{j,k\neq i}q_jq_k
    \frac{\boldsymbol{r}_{ij}\cdot \boldsymbol{r}_{ik}}{|\boldsymbol{r}_{ij}|^2|\boldsymbol{r}_{ik}|^2},
\end{align}
where we have defined $\boldsymbol{r}_{ij} = \boldsymbol{r}_i - \boldsymbol{r}_j$. We shall restrict ourselves to two-anyon systems, and leave more complicated scenarios for future study. We specifically focus on a bound state of a quasihole and a quasielectron -- the fractional exciton. The Hamiltonian for this system is the following 

\begin{equation}\label{eq:Haa:expanded}
    H^{aa} = - \frac{\nabla_1^2 + \nabla_2^2}{2m_a}
    - i \frac{\theta}{m_a} \frac{\left[ \boldsymbol{r}_{12} \times \left( \nabla_1 - \nabla_2 \right) \right]_z}{|\boldsymbol{r}_{12}|^2}
    + \frac{\theta^2}{m_a} \frac{1}{|\boldsymbol{r}_{12}|^2} + V^\text{KR}(|\boldsymbol{r}_{12}|)
\end{equation}

Introducing COM coordinates
\begin{equation}
    \boldsymbol{R} = \frac{\boldsymbol{r}_1 + \boldsymbol{r}_2}{2}, \qquad \boldsymbol{\rho} = \boldsymbol{r}_1-\boldsymbol{r}_2
\end{equation}
the Hamiltonian simplifies to

\begin{equation}
    H^{aa} = - \frac{\nabla_R^2}{4m_a} - \frac{\nabla_\rho^2}{m_a}
    - i \frac{\theta}{m_a} \frac{2(\boldsymbol{\rho}\times\nabla_{\boldsymbol{\rho}})_z}{|\boldsymbol{\rho}|^2}
    + \frac{\theta^2}{m_a} \frac{1}{|\boldsymbol{\rho}|^2} + V^\text{KR}(|\boldsymbol{\rho}|).
\end{equation}

Note that $-i(\boldsymbol{\rho}\times \nabla_{\boldsymbol{\rho}})_z=L_z$ is the $z$-component angular momentum about $\boldsymbol{\rho}$, so we shall generalize the ECG basis by including functions with angular momentum
\begin{equation}
    \sum_{l\ge 0} (\rho_x+ i\rho_y)^l \phi_\alpha(\boldsymbol{R},\boldsymbol{\rho}), \qquad \sum_{l < 0} (\rho_x- i\rho_y)^l \phi_\alpha(\boldsymbol{R},\boldsymbol{\rho})\,.
\end{equation}
The statistical interaction is singular, and we regulate it by smearing at the Moir\'{e} scale $\ell_0$
\begin{equation}
    \frac{1}{|\boldsymbol{\rho}|^2} \rightarrow \frac{1}{|\boldsymbol{\rho}|^2 + \ell_0^2}
\end{equation}
where $\ell_0\approx 6$ nm for $t$MoTe$_2$ at twist angle $\approx 3.7^\circ$~\cite{cai2023signatures,redekop2024direct}.

\begin{figure}[t]
    \centering
    \includegraphics[width=0.5\columnwidth]{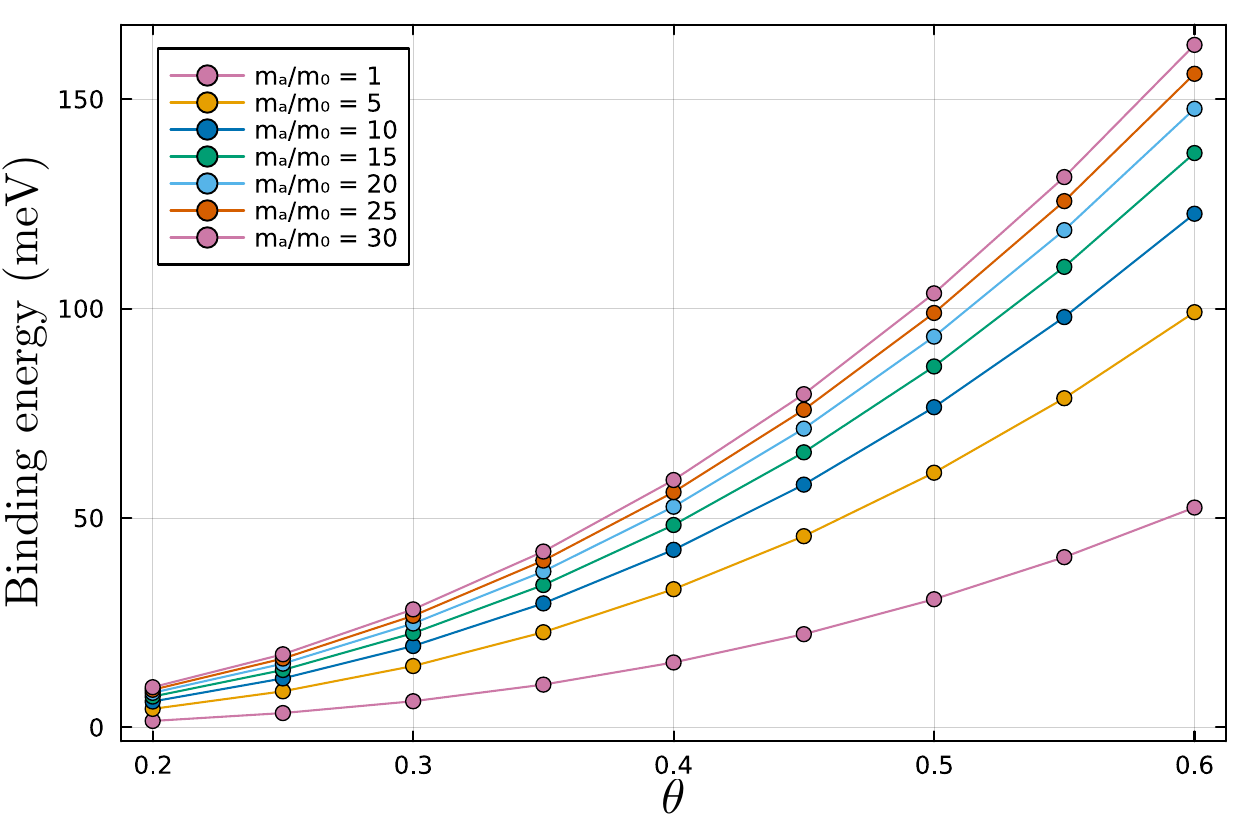}
    \caption{Binding energy of fractional exciton made by an anyon and an anti-anyon.}
    \label{fig:fractional_exciton_energy}
\end{figure}

For the numerics, we restrict ourselves to $|l|\le 1$. With the above setup, the ground state energy of the fractional exciton for various masses and charges is plotted in Fig.~\ref{fig:fractional_exciton_energy}

\FloatBarrier
\onecolumngrid 
\section{Additional numerical data}\label{app: AdditionalData}

\subsection{Fractionally charged trions: Anyon-exciton bound states}

\begin{figure}[t]
    \centering
    \includegraphics[width=0.5\columnwidth]{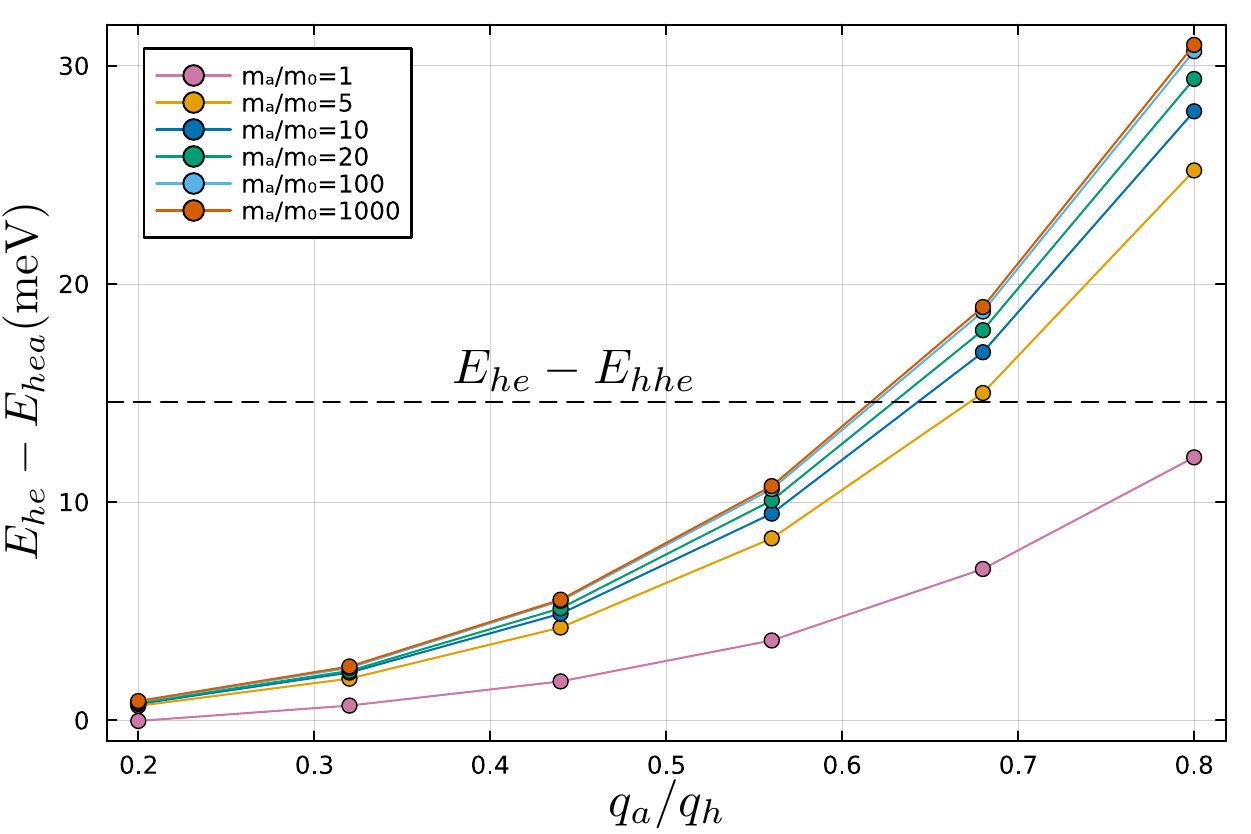}
    \caption{Binding energy for an anyon to an exciton, $E_{he}-E_{hea}$ (without a trap) as a function of anyon mass $m_a$ and charge $q_a$. The dashed horizontal line marks the binding energy of the ordinary trion $E_{he}-E_{hhe}$. With $q_a$ increasing, the linear approximation of the binding energy is no longer valid. We use $m_h=m_e=0.65m_0$.}
    \label{fig: Ehha binding energy}
\end{figure}

In the main text, we mentioned the possibility that in the absence of a trap an exciton can bind an anyon instead of a hole, forming a \emph{fractionally charged trion} ($T_\star$). These objects are ground states of the three-body problem of one electron, one hole, and one anyon. 

Fig.~\ref{fig: Ehha binding energy} shows the fractional trion binding energy
\begin{equation}
    \delta E_{T_\star}^{\rm{free}} = E_{he}^{\rm{free}}-E_{hea}^{\rm{free}},
\end{equation}
and compares it with the tradional trion's binding energy $\delta E_T^{\rm{free}} = E_{he}^{\rm{free}}-E_{hhe}^{\rm{free}}$. For large enough $m_a$ and $q_a$, the fractionally charged trion has a larger binding energy than the traditional trion, so that an anyon will replace a hole in a trion upon collision.

The anyon-exciton $hea$ is in general not favored energetically, especially for $q_a/q_e<2/3$. Nevertheless, as the anyon mass and charge increases, the binding energy grows and may exceed $E_T$. This result suggests that if the itinerant charge carriers are multi-anyon bound states with very narrow bandwidth, an additional fractional trion feature may become visible and overtake the free trion. In the following, we plot $E_{hea}-E_{he}$ and compare it with $E_T=E_{hhe}-E_{he}$. An important implication from Fig.~\ref{fig: Ehha binding energy} is that the fractional trion can appear in the spectrum if the anyon is heavy enough and $q_a/q_e$ is approximately larger than $2/3$. Nevertheless, the binding energy in the large $q_a$ region cannot be simply approximated by a linear function of $q_a$, which make complicate the experimental observation, compared with the anyon-trion at a small $q_a$ discussed in the main text.
\begin{figure*}[t]
    \centering
    \includegraphics[width=0.48\textwidth]{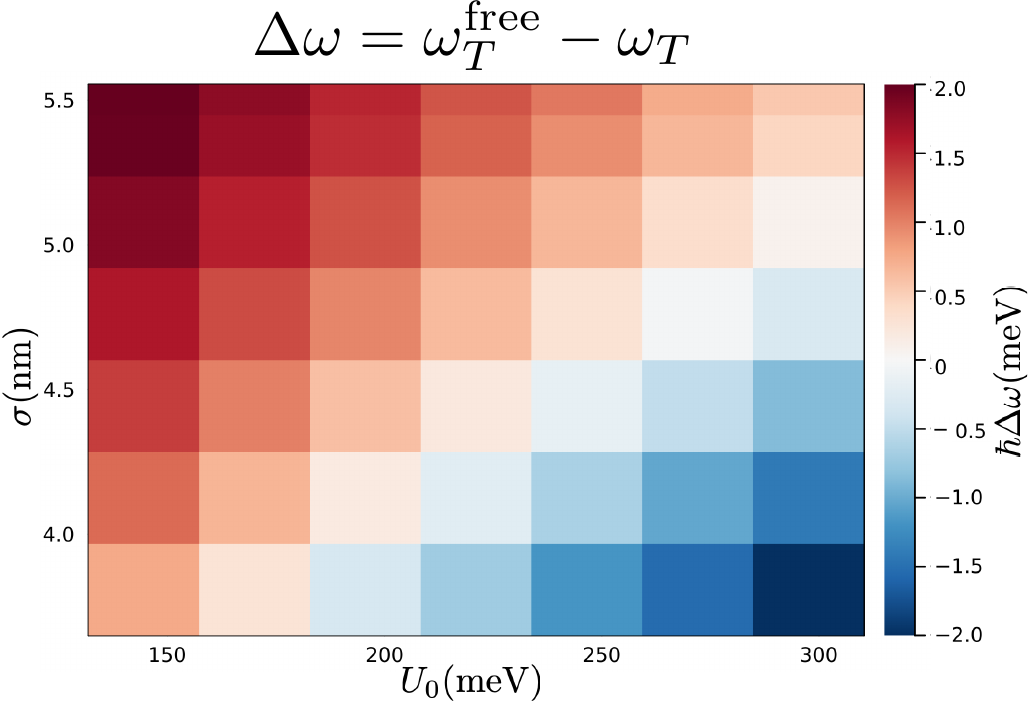}\hfill
    \includegraphics[width=0.48\textwidth]{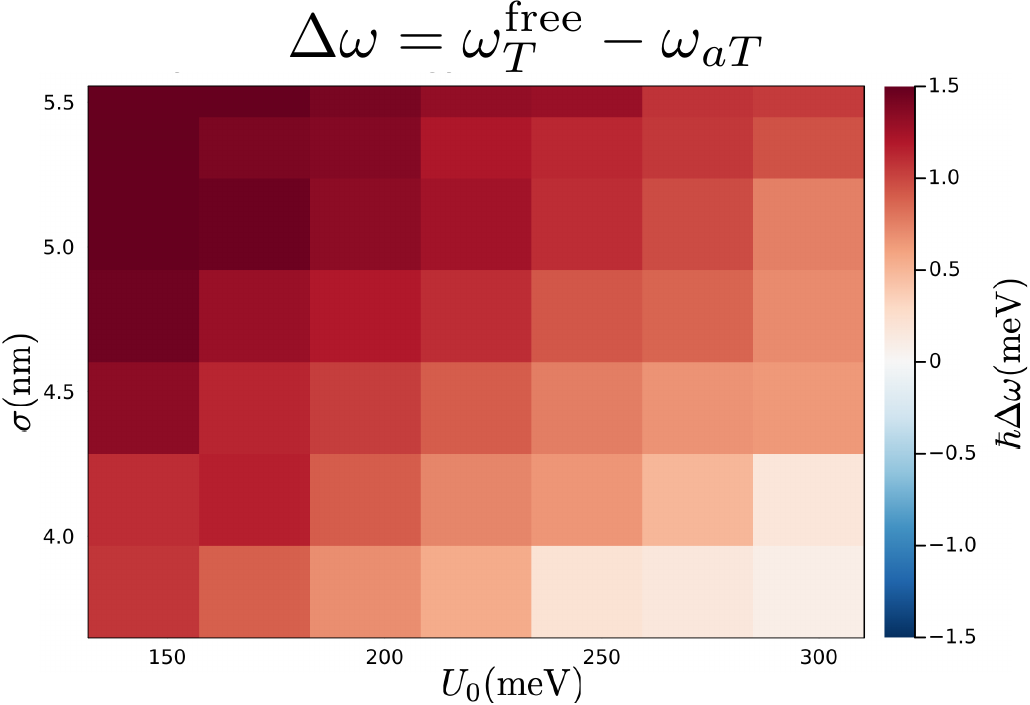}
       \caption{\textit{Left panel:} Trapped trion energy relative to free trion. \textit{Right panel:} Anyon-trion energy relative to free trion. 
       We have used an anyon charge $q_a=1/3$ and anyon mass $m_a = 10\, m_0$. The difference between the left and the right panel gives Fig.~\ref{fig:V0_vs_sigma_stability}
       }.
    \label{fig:FrequencyShiftFromFreeTrion}
\end{figure*}

\FloatBarrier
\subsection{Trion energies}

In Fig.~\ref{fig:V0_vs_sigma_stability} of the main text, we show the anyon-induced red-shift from the trapped trion:
\begin{equation}
    \hbar \Delta \omega_a =
    \hbar (\omega_{T}-\omega_{aT}) =
    (E_T-E_h)-(E_{aT}-E_{ah}).
\end{equation}
We focused on this difference because it directly encodes the presence of a valence anyon. 

Ref.~\cite{li2026signatures} also reported that the trap red-shifts the trion peak, {i.e.} $\hbar(\omega_{T}^{\rm{free}}-\omega_{T} )>0$.
In our model,
\begin{equation}
     \hbar(\omega_{T}^{\rm{free}}-\omega_{T} ) = 
     E_{T}^{\rm{free}}-(E_{T} - E_{h}),
\end{equation}
can be of either sign depending on the trap parameters. In
Fig.~\ref{fig:FrequencyShiftFromFreeTrion}, we show this energy difference, as well as, $\hbar(\omega_{T}^{\rm{free}} -\omega_{aT}) = 
    (E_{ah}-E_{aT})-(-E_T^{\rm{free}})$. We see that the trapped trion PL peak can be blue-shifted ($\omega_{T}^{\rm{free}}-\omega_{T}<0$) with respect to the free trion if the trap is too shallow too wide.

\FloatBarrier
\subsection{Spatial structure of the bound anyonic molecule}

\begin{figure*}[t]
    \centering
    \includegraphics[width=0.48\textwidth]{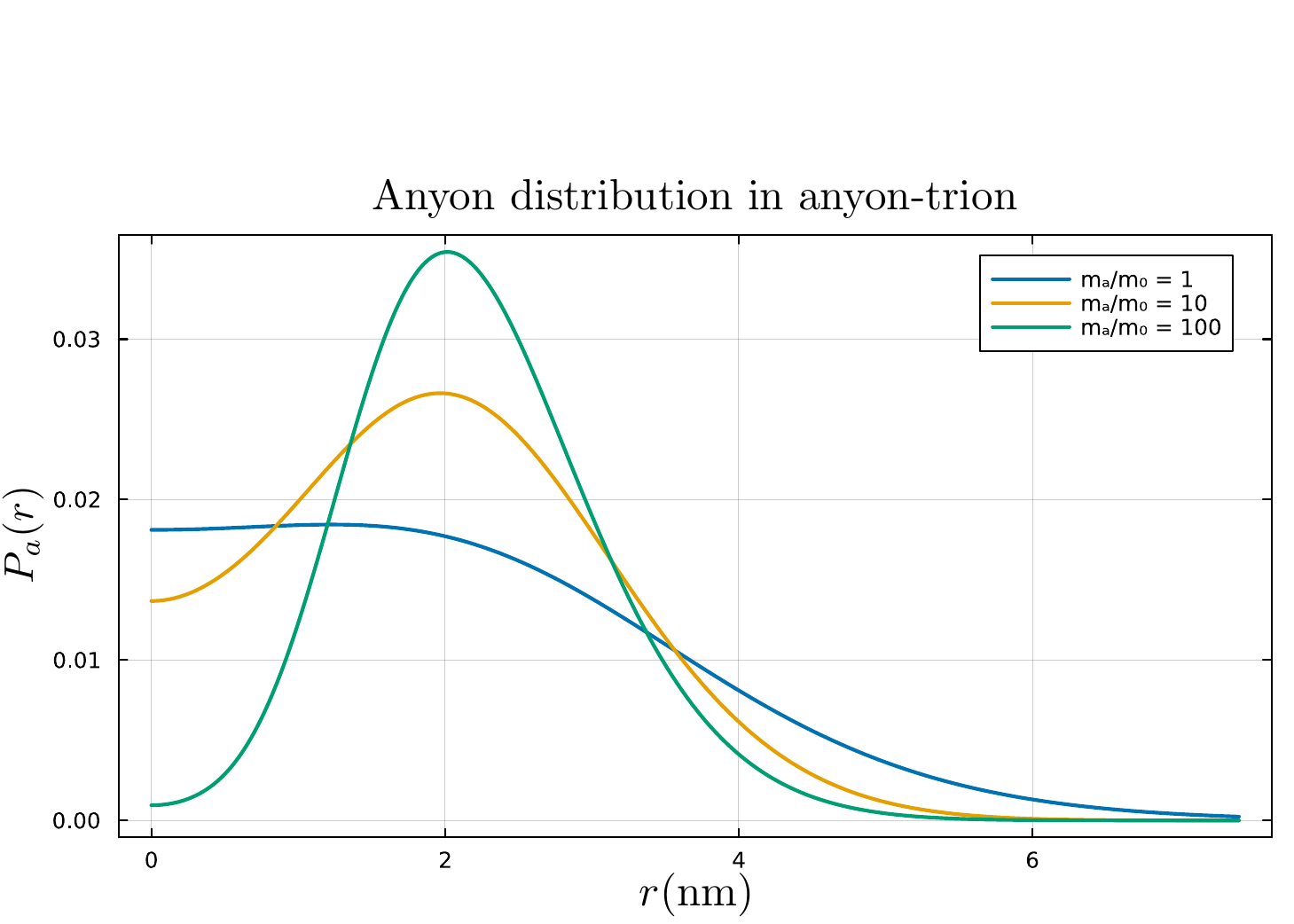}\hfill
    \includegraphics[width=0.48\textwidth]{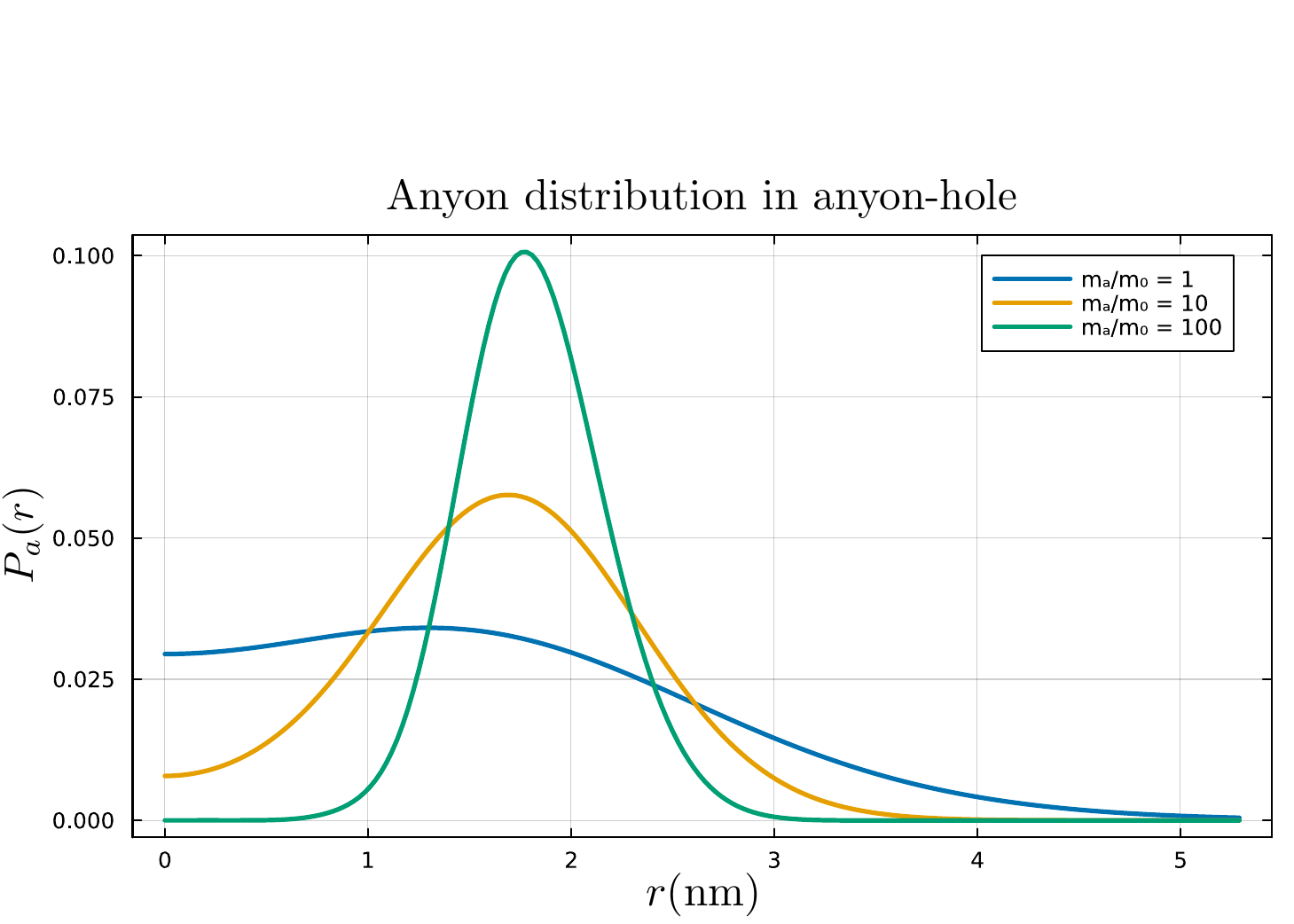}
    \caption{Anyon density $\mathrm{P}_a(r)$ of anyon-trion (left panel) and anyon-hole (right panel) system in the presence of the trap as a function of its distance to the trap bottom ($r$) for several valus of anyon mass $m_a$ and $q_a=1/3$. The trap parameter {$\sigma=4.2$nm} and {$U_0=190$meV}.
    }
    \label{fig:anyon density profile}
\end{figure*}

\begin{figure*}[t]
    \centering
    \includegraphics[width=0.48\textwidth]{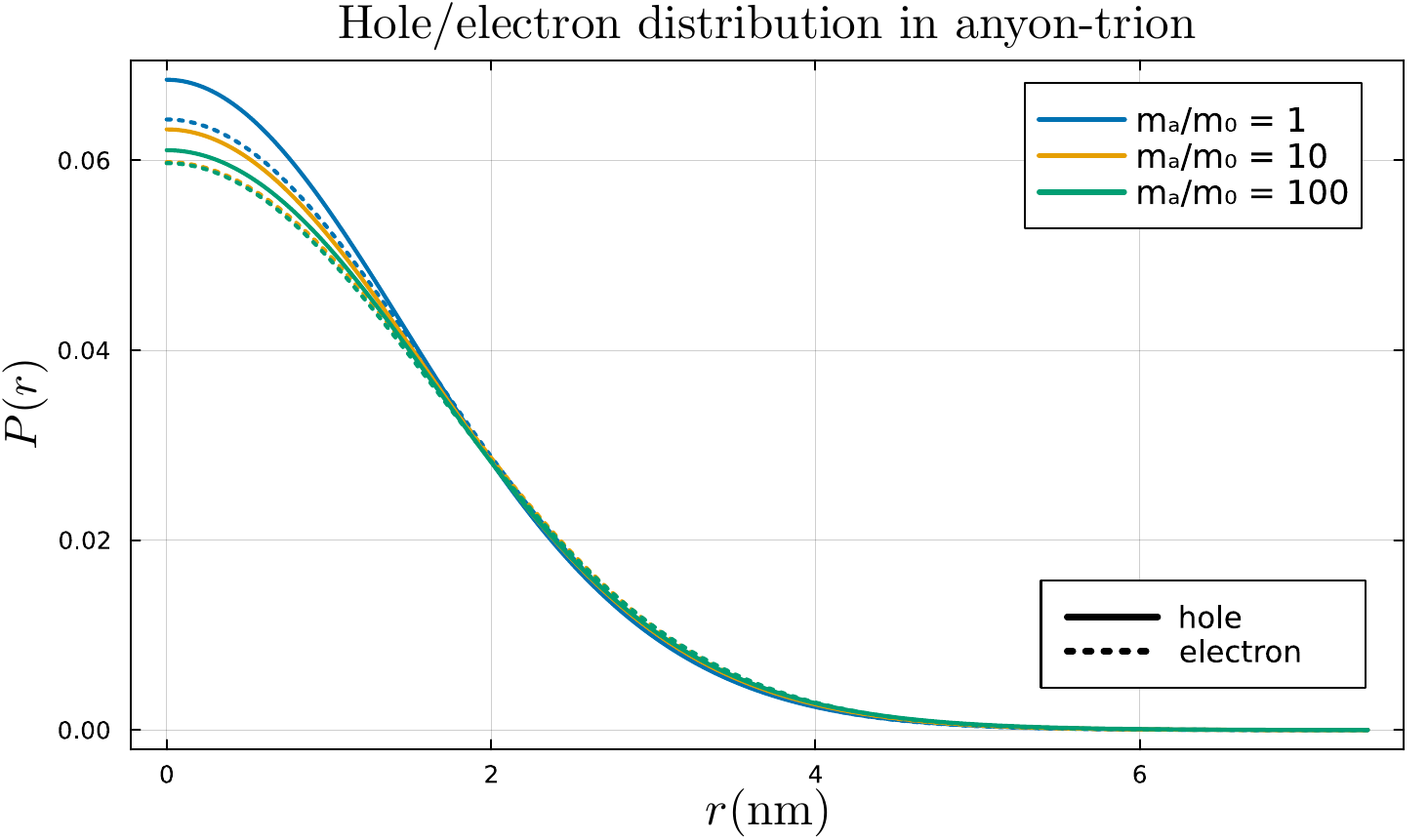}\hfill
    \includegraphics[width=0.48\textwidth]{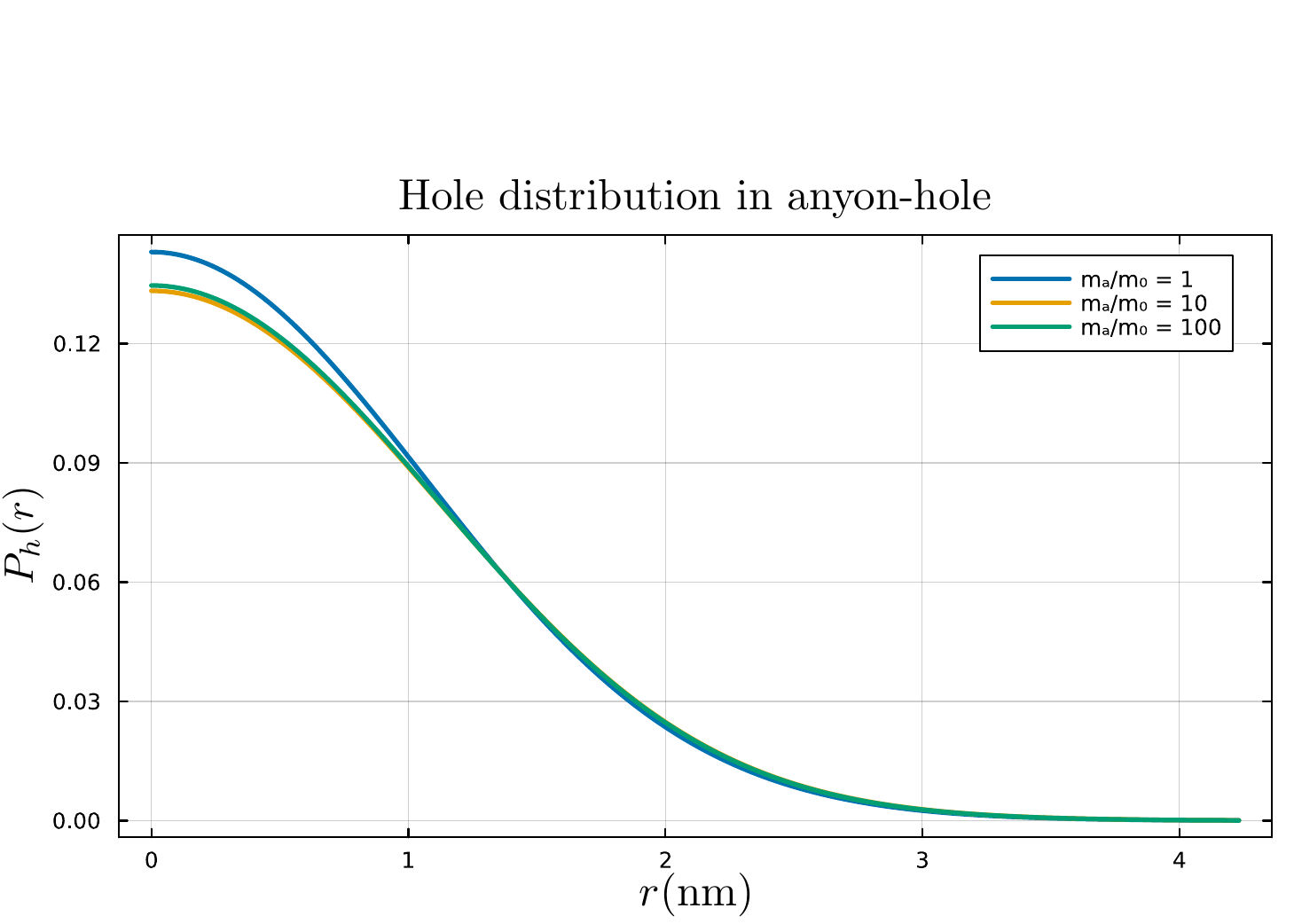}
    \caption{Hole and electron density $\mathrm{P}_h(r), \mathrm{P}_e(r)$ of anyon-trion (left panel) system, and hole density $\mathrm{P}_h(r)$ of anyon-hole (right panel) system  in the presence of the trap as a function of its distance to the trap bottom ($r$) for several valus of anyon mass $m_a$ and $q_a=1/3$. The trap parameter $\sigma=4.2$ nm and $U_0=190$ meV.  }
    \label{fig:all particle density profile}
\end{figure*}

Define the density function of particle $i$ around the trap center by
\begin{align}
    \mathrm{P}_i(\boldsymbol{r}_i)=\int\, \prod_{j \neq i} \abs{\psi}^2 d^2\boldsymbol{r}_{j},
\end{align} 
where $\psi$ is the variational ground state wavefunction, and $\boldsymbol{r}_i$ is measured from the center of the trap. The density function is normalized such that $\int \mathrm{P}(\boldsymbol{r}_i)\, d^2\boldsymbol{r}_i=1$.

In Fig.~\ref{fig:anyon density profile}, we show $\mathrm{P}_i(\boldsymbol{r}_i)$ for the anyon in the presence of either a trion or hole. 
By rotational symmetry of the trial state, $\mathrm{P}(\boldsymbol{r}_i)=\mathrm{P}(r_i)$ where $r_i=\abs{\boldsymbol{r}_i}$. 
We see that for sufficiently large mass the anyon tends to localize near a particular value of $r_a$, which depends on the particle at the bottom of the trap. 

In Fig.~\ref{fig:all particle density profile}, we show $\mathrm{P}_i(\boldsymbol{r}_i)$ for the ordinary particles (electron and/or hole) within the anyon-trion composite and anyon-hole composite, respectively. Their distribution is not very sensitive to the mass of anyon.

\end{document}